# Engineering Nonporous Polymer Hybrids with Suppressed Heat Conduction and Enhanced Flame Retardancy via Molecular and Filler Design

Henry Worden,[1#] Mihir Chandra,[2#] Yijie Zhou,[1#] Zarif Ahmad Razin Bhuiyan,[3] Mouyang Cheng,[4] Krishnamurthy Munusamy,[5] Weiguo Hu,[5] Weibo Yan,[1] Siyu Wu,[6] Ruipeng Li,[6] Anna Chatterji,[1] Todd Emrick,[5] Jun Liu,[3] Yanfei Xu[1,2*]

1. Department of Mechanical and Industrial Engineering, University of Massachusetts, Amherst, Massachusetts, 01003, United States of America
2. Department of Material Science and Engineering, University of Massachusetts, Amherst, Massachusetts, 01003, United States of America
3. Department of Mechanical and Aerospace Engineering, North Carolina State University, Raleigh, NC 27695, USA.
4. Department of Materials Science and Engineering, Massachusetts Institute of Technology, Cambridge, MA 02139, USA.
5. Department of Polymer Science and Engineering, University of Massachusetts, Amherst, Massachusetts, 01003, United States of America
6. Brookhaven National Laboratory, Upton, New York, 11973, United States of America

#Contribute equally

*Corresponding author yanfeixu@umass.edu

**Table of Contents**

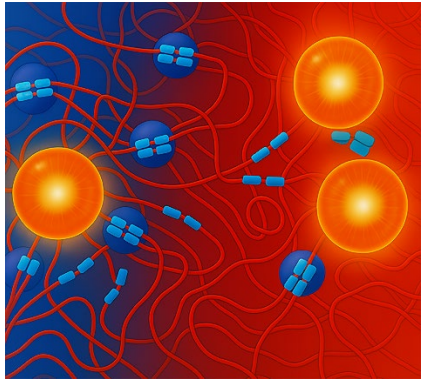

This study presents a new strategy for achieving ultralow thermal conductivity in nonporous polymer/organic filler hybrids by suppressing heat capacity through tailored atomic vibrations to enhance thermal insulation. Unlike conventional polymer/inorganic filler hybrids, these hybrids exhibit interfacial thermal resistance one to three orders of magnitude lower. Combined experiments and simulations uncover thermal transport mechanisms. These hybrids demonstrate enhanced flame retardancy.

**Abstract:**

Achieving ultralow thermal conductivity and enhanced flame retardancy in polymers, while deepening the fundamental understanding of thermal transport mechanisms, remains a long-standing challenge. Although interfacial thermal resistance in polymer/inorganic fillers is commonly thought to hinder heat conduction, our results demonstrate that interfacial thermal resistance in polymer/organic filler (PU/THDBT) hybrids is negligible, exhibiting resistances of ~$10^{-9}$ $m^2$ K $W^{-1}$—one to three orders of magnitude lower. We demonstrate for the first time that suppressed heat capacity in the PU/THDBT hybrids—by tailoring atomic vibrations through controlled chemical compositions and structural design—enables more effective reduction of thermal conductivity. Specifically, PU/THDBT hybrids containing ~23 vol% THDBT exhibited a ~17% reduction in thermal conductivity (from ~0.141 W $m^{-1}$ $K^{-1}$ to ~0.117 W $m^{-1}$ $K^{-1}$), relative to pure PU. Moreover, incorporation of THDBT fillers significantly enhances the flame retardancy of PU/THDBT hybrids. These findings establish heat-capacity suppression as a fundamentally new design principle for nonporous polymer/organic filler hybrids with ultralow thermal conductivity, complementing traditional approaches that rely on engineering interfacial thermal resistance via heterogeneous domain structures, and advancing the understanding of thermal transport in polymer hybrids.

**Introduction**

Polymers uniquely combine light weight, chemical resistance, and ease of processing—properties that are lacking in traditional metals and ceramics.[1] These advantages have made polymers attractive for thermal insulation across healthcare, construction, aerospace, and related sectors.[2-11] Because of their inherently low thermal conductivities (on the order of ~0.1 W m$^{-1}$ K$^{-1}$),[1,12] polymers are well known as effective thermal insulators and promising candidates for thermal insulation management.[13-16] Pushing polymer thermal conductivities even lower (e.g., toward ~0.03 W m$^{-1}$ K$^{-1}$) is especially compelling, given that porous foams or aerogels—among the best-known thermal insulators—exhibit values of ~$0.02 - 0.04$ W m$^{-1}$ K$^{-1}$.[5,9,17-24]

Common approaches to engineering polymer with ultralow thermal conductivity rely on porous or hollow structures to suppress heat conduction.[3,17,21,25] However, nonporous polymers with ultralow thermal conductivities are particularly desirable in extreme thermal environments.[26] They provide superior compressive resistance to satisfy mechanical-strength requirements and also serve as effective barriers against oxygen and moisture transport.[27-29] In inorganic hybrids such as alloys, thermal conductivity can be reduced without introducing porous structures by enhancing interfacial thermal resistance through interface engineering.[30,31] In polymer hybrids, however, this strategy is considerably less effective because conventional inorganic fillers, such as metals and ceramics, possess intrinsic thermal conductivities far higher than those of polymers, leading to overall high thermal conductivities in polymer/metal or polymer/ceramic hybrids.[32-35] Therefore, achieving ultralow thermal conductivity in nonporous polymers remains a significant challenge.[5,12] This challenge arises from the limited understanding of heat transport in polymers, in part due to the insufficient knowledge of how chemical compositions and structures govern thermal transport properties. While adding inorganic metallic or ceramic fillers is not a viable route, incorporating organic fillers offers a promising strategy to tune polymer hybrid chemical compositions and structures and to probe thermal transport mechanisms in polymers. To date, however, experimental research in understanding thermal transport in polymer/organic filler hybrids has been limited. Addressing these knowledge gaps is essential to establish design rules for nonporous polymers with ultralow thermal conductivities that rival those of state-of-the-art thermal insulators.

For the first time in the experimental field of thermal transport in nonporous polymer/organic filler hybrids, this work provides experimental evidence that tailoring chemical compositions and structures in these hybrids can effectively reduce heat capacity and suppress heat conduction. Heat conduction is reduced through atomic vibrational engineering, which lowers the effective specific heat capacity. As a model system of nonporous polymer/organic filler hybrids, we use polyurethane (PU) as the polymer (block copolymer) matrix and tetrahydroxy deoxybenzoins triazole (THDBT) as the organic filler. The PU/THDBT hybrid containing ~23 vol% THDBT exhibited a ~17% reduction in thermal conductivity compared to pure PU.

Contrary to the conventional view,[36-38] our experiments and molecular dynamics (MD) simulations demonstrate that polymer/organic filler interfaces (PU/THDBT) do not significantly hinder thermal transport. The interfacial thermal resistance between PU and THDBT is negligible, on the order of $10^{-9}\ m^2\ K\ W^{-1}$—one to three orders lower than polymer/inorganic filler interfaces ($10^{-6}$–$10^{-8}\ m^2\ K\ W^{-1}$).[12,30,39-43] These findings reveal distinct interfacial behavior in PU/THDBT hybrids, challenging the assumption that heterogeneous interfaces inherently suppress heat transport.

In addition to polymer/organic filler hybrids with ultralow thermal conductivity for thermal insulation, flame retardancy is another critical property desired for thermal management applications.[44,45] This is because the carbon- and hydrogen-rich chemistry of polymers renders them inherently flammable,[44,45] creating significant risks in high-temperature environments. However, the design of polymers that simultaneously deliver ultralow thermal conductivity and enhanced flame-retardant performance remains limited.[45] Two key challenges for designing underpin this limitation: (1) Although incorporating flame-retardant fillers into polymer matrices is a widely adopted strategy, the flame retardance effectiveness of these fillers depends critically on their dispersion, chemical compatibility, and interfacial interactions—factors that remain insufficiently understood.[46,47] (2) Suppressing thermal conductivity in nonporous polymers is complicated by their inherent structural disorders, as well as the limited understanding of how chemical compositions and chain structures govern thermal transport properties in polymer/organic filler hybrids.[7,8,32] Addressing these challenges requires systematic experimental studies to establish design principles for polymer/organic filler hybrids capable of simultaneously achieving ultralow thermal conductivity and robust flame-retardant performance.

Here, we show for the first time that organic fillers can be engineered to reduce specific heat capacity, lower thermal conductivity, and enhance flame retardancy in nonporous polymer/organic filler hybrids. THDBT fillers are uniformly dispersed within the PU matrix through hydrogen bonding between functional groups on the THDBT fillers and the PU. These fillers improve flame retardancy by promoting char formation during combustion. The PU/THDBT hybrid containing ~23 vol% THDBT exhibits a 27% reduction in fire growth capacity compared to pure PU.

We use block copolymer systems as model platforms to design nonporous polymer/organic filler hybrids with suppressed thermal conductivity and enhanced flame retardancy while elucidating the underlying thermal transport mechanisms (**Figure 1**). This choice is motivated by two factors: (1) Technological significance—block copolymers such as PU are widely used as model polymer systems in building insulation and thermal management applications;[11,48] and (2) Fundamental interest—block copolymers, such as PU, are of fundamental interest due to their structures comprising alternating soft (flexible) and hard (rigid) segments, which typically phase-separate due to segmental incompatibility (**Figures 1A and 1B**).[49-52] The tunable structures and compositions of the soft and hard segments enable the engineering of phase-separated morphologies with nano-domain sizes, as well as the manipulation of intra- and intermolecular interactions. This provides a rich platform for studying how nanoscale phase structures—typically comprising domains on the order of tens of nanometers—affect thermal transport properties, including whether interfacial regions serve as bottlenecks for heat conduction.

We use THDBT as a model organic filler to design nonporous polymer/organic filler hybrids that simultaneously suppress heat conduction and enhance flame retardancy (**Figures 1C and 1D**). This approach is motivated by two factors: (1) The incorporation of functional groups (e.g., the benzene rings and triazole rings) in amorphous THDBT provides a model platform to investigate how atomic vibrations influence specific heat capacity, as the role of atomic vibrations in governing specific heat remains poorly understood in amorphous organic molecules and polymers. (2) these designed functional groups provide a platform to investigate how flame retardancy in polymer/organic filler hybrids can be enhanced through the promotion of char formation.

Thus, PU/THDBT hybrids provide a model platform for elucidating how chemical compositions and structures govern thermal transport and flame-retardant behavior in nonporous polymer/organic filler hybrids. The synthesis details of the PU/THDBT hybrids are provided in Section S1 in the Supplementary Materials.

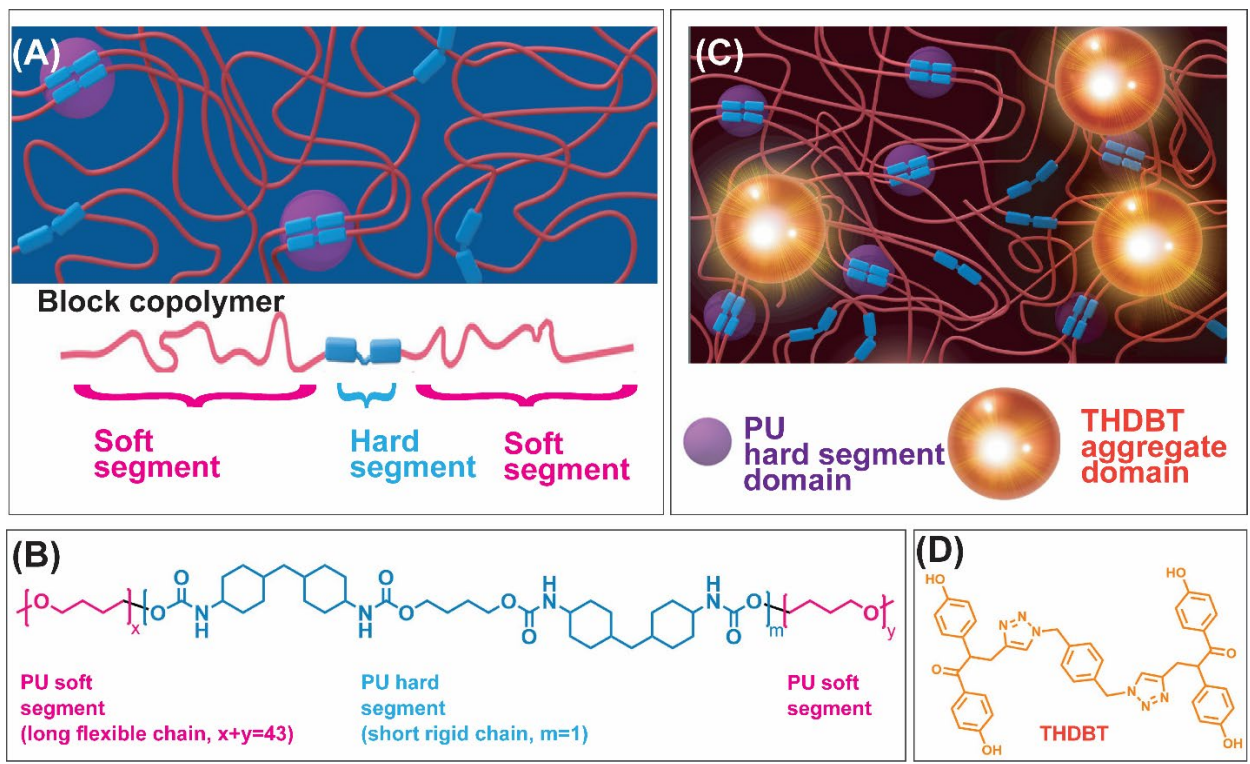

**Figure 1.** Study design for investigating thermal transport mechanisms in nonporous polymer/organic filler hybrids. **(A)** A schematic illustration of block copolymer chains, in which the polymer backbones have alternating soft (flexible) segments and hard (rigid) segments. The soft segments are long, flexible, and entangled chains, while the hard segments are short and rigid. **(B)** Chemical structures of PU with hard and soft segments. **(C)** A schematic illustration of polymer/organic filler (PU/THDBT) hybrid structure. **(D)** Chemical structures of THDBT containing benzene rings and triazole rings.

### Results and Discussion

Experimentally measuring cross-plane thermal diffusivity and thermal conductivity is essential for understanding and optimizing the thermal insulation performance of polymer hybrids such as nonporous PU/THDBT hybrids, particularly for applications requiring resistance to heat conduction perpendicular to the surface. We calculate the cross-plane thermal conductivity ($k$, $W\,m^{-1}\,K^{-1}$) using **Equation 1** for the solids studied in this work. **Equation 1** can be rewritten as **Equation 2** using volumetric heat capacity as a calculation parameter.[43,53,54] These solids include pure PU films, compressed pellets of THDBT fillers, and PU/THDBT hybrids containing THDBT fillers at various volume fractions of 4.5 vol%, 13.9 vol%, and 23.3 vol%. For clarity, these PU/THDBT hybrids are denoted as PU/THDBT (4.5 vol%), PU/THDBT (13.9 vol%), and PU/THDBT (23.3 vol%). We note that both PU and THDBT are electrically insulating; therefore, heat conduction in PU, compressed pellets of THDBT fillers, and PU/THDBT hybrids are dominated by atomic vibrations, with negligible electronic contributions.

$$k = \alpha \times c_p \times \rho \quad (Equation\ 1)$$

$$k = \alpha \times c_v \quad (Equation\ 2)$$

We measure thermal diffusivity ($\alpha$, m$^2$ s$^{-1}$) in the cross-plane direction using the laser flash technique.[43] The specific heat capacity ($c_p$, J g$^{-1}$ K$^{-1}$) is determined by the differential scanning calorimetry technique. The $c_v$ is the volumetric heat capacity (J K$^{-1}$ m$^{-3}$). The density ($\rho$, g cm$^{-3}$) is measured using Archimedes' method. The cross-plane thermal conductivity ($k$) is obtained using **Equation 1**. We assume this cross-plane $k$ value represents the thermal conductivity in all dimensions due to isotropic structures of PU/THDBT hybrids, as characterized by X-ray scattering analysis (**Figure S1** in the Supplementary Material).

The measured thermal diffusivity values ($\alpha$) are presented in **Figure 2A**. A measured thermal diffusivity of pure PU films is ~0.078 ± 0.005 mm$^2$ s$^{-1}$, while compressed pellets of THDBT fillers exhibit a value of ~0.117 ± 0.003 mm$^2$ s$^{-1}$. The measured thermal diffusivities are ~0.074 ± 0.001 mm$^2$ s$^{-1}$, ~0.078 ± 0.005 mm$^2$ s$^{-1}$, and ~0.077 ± 0.004 mm$^2$ s$^{-1}$ for PU/THDBT hybrids (4.5 vol%), PU/THDBT hybrids (13.9 vol%), and PU/THDBT hybrids (23.3 vol%), respectively (**Figure 2A**, **Figures S2–S6**).

The measured specific heat capacities ($c_p$) are presented in **Figure 2B**. The specific heat capacity of the compressed pellets of THDBT fillers (~0.749 ± 0.059 J g$^{-1}$ K$^{-1}$) is markedly lower than the pure PU films (~1.671 ± 0.083 J g$^{-1}$ K$^{-1}$). The measured specific heat capacities are ~1.628 ± 0.021 J g$^{-1}$ K$^{-1}$, ~1.522 ± 0.013 J g$^{-1}$ K$^{-1}$, and ~1.373 ± 0.037 J g$^{-1}$ K$^{-1}$ for PU/THDBT hybrids (4.5 vol%), PU/THDBT hybrids (13.9 vol%), and PU/THDBT hybrids (23.3 vol%), respectively (**Figure 2B**). Compared with pure PU films, the experimentally measured low specific heat capacities in the compressed pellets of THDBT fillers and PU/THDBT hybrids originate from the engineered atomic vibrations within these materials. This observation validates our design principle that the rigid bonds in the functional groups (e.g., benzene and triazole rings) of THDBT restrict vibrational freedom, leading to low specific heat capacity.

The measured density values ($\rho$) are presented in **Figure 2C**. Pure PU films exhibit a measured density of ~1.101 ± 0.023 g cm$^{-3}$, while compressed pellets of THDBT fillers show a value of ~1.144 ± 0.016 g cm$^{-3}$. The measured density are ~1.094 ± 0.018 g cm$^{-3}$, ~1.100 ± 0.014 g cm$^{-3}$, and ~1.115 ± 0.011 g cm$^{-3}$ for PU/THDBT hybrids (4.5 vol%), PU/THDBT hybrids (13.9 vol%), and PU/THDBT hybrids (23.3 vol%), respectively (**Figure 2C**).

The measured thermal conductivity values ($k$) are presented in **Figure 2D**. Pure PU films exhibit a measured thermal conductivity of ~0.141 ± 0.011 W·m$^{-1}$ K$^{-1}$, while compressed pellets of THDBT fillers show a value of ~0.090 ± 0.007 W m$^{-1}$ K$^{-1}$. The measured thermal conductivities are ~0.131 ± 0.003 W m$^{-1}$ K$^{-1}$, ~0.130 ± 0.010 W m$^{-1}$ K$^{-1}$, and ~0.117 ± 0.007 W m$^{-1}$ K$^{-1}$ for PU/THDBT hybrids (4.5 vol%), PU/THDBT hybrids (13.9 vol%), and PU/THDBT hybrids (23.3 vol%), respectively (**Figure 2D**). Because thermal conductivity depends on specific and volumetric heat capacities, lowering the specific heat capacity provides an effective strategy to suppress thermal conductivity in polymer/filler hybrids.

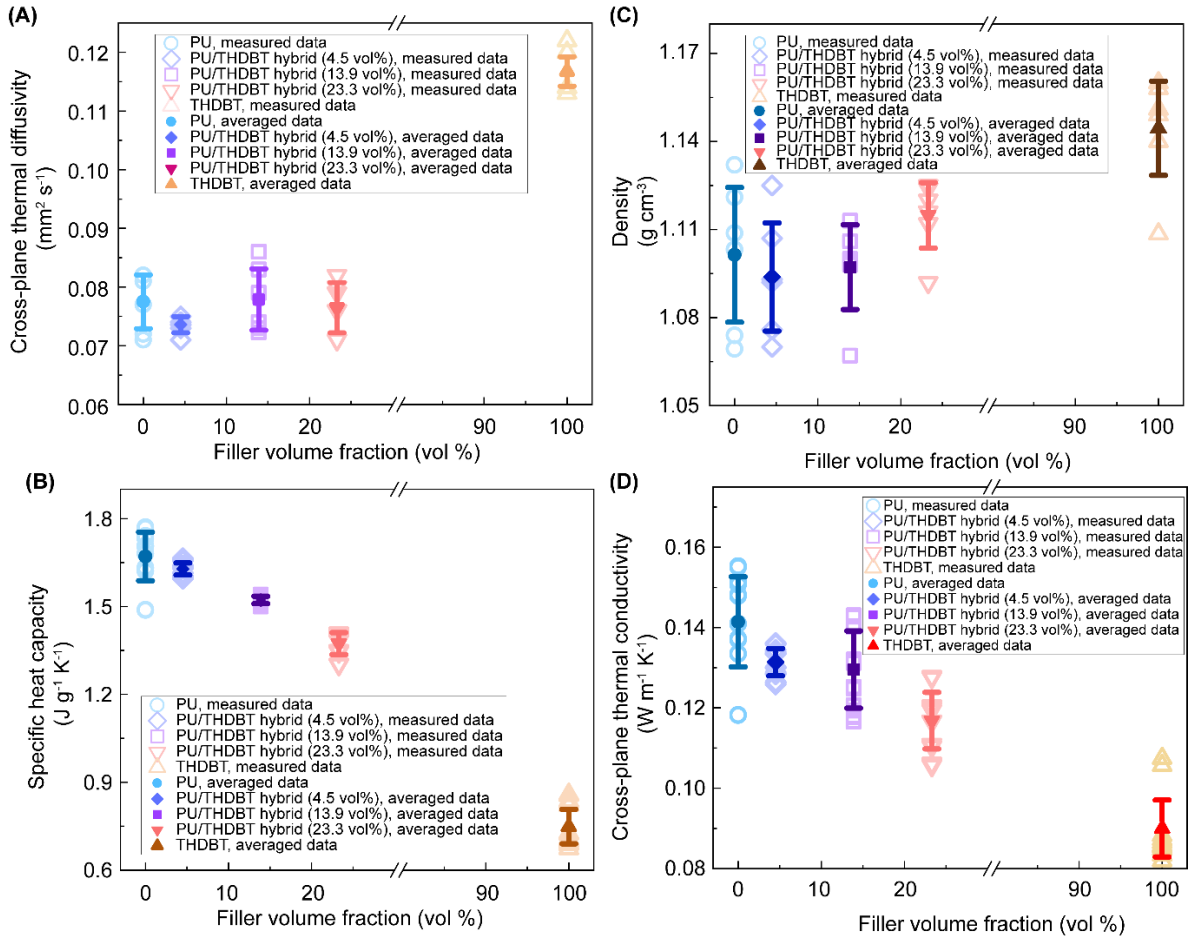

**Figure 2.** Experimental study of thermal transport mechanisms in polymers, organic fillers, and nonporous polymer/organic filler hybrids. **(A)** Cross-plane thermal diffusivities measured at 298 K. The samples include pure PU thin film, PU/THDBT (4.5 vol%), PU/THDBT (13.9 vol%), PU/THDBT (23.3 vol%), and compressed pellets of THDBT. Error bars of thermal diffusivity measurements primarily caused by thickness differences of the polymer hybrids. The error bars in the thermal diffusivity measurements primarily arise from variations in the sample thickness and the differences in thermal diffusivity observed across six measurements for each sample.[43] **(B)** Specific heat capacities measured at 298 K. The error bars for the specific heat capacities represent the population standard deviation from three measurements for each sample.[43] **(C)** Densities measured at 298 K. Error bars, calculated as the population standard deviation from six measurements per sample, primarily arise from variations in volume measurements.[43] **(D)** Cross-plane thermal conductivities measured at 298 K. Error bars of thermal conductivities calculated based on error propagations of cross-plane thermal diffusivity, specific heat capacity, and density measurements of PU hybrids.[43] Details of the error analysis are provided in Section S2 in the Supplementary Material.

To determine whether the interfacial thermal resistance ($R_{ITR}$) between the PU polymer and THDBT filler significantly impedes thermal transport and reduces the measured thermal conductivity of PU/THDBT hybrids at the atomic level, we determine $R_{ITR}$ at PU/THDBT interfaces using molecular dynamics (MD) simulations.[55-59] The $R_{ITR}$ at the two interfaces are $\sim 0.5 \times 10^{-9}$ m$^2$ K W$^{-1}$ and $\sim 1.1 \times 10^{-9}$ m$^2$ K W$^{-1}$ respectively, corresponding to a 1.49 K and 2.97 K temperature drop at the first and second interfaces (**Figures 3A and 3B**, Section S3 in the Supplementary Materials). Average $R_{ITR}$ is ~$0.80 \times 10^{-9}$ m$^2$ K W$^{-1}$.

In contrast to the conventional view that interface engineering suppresses heat conduction by increasing interfacial thermal resistance in polymer/inorganic filler hybrids, our results show that polymer/organic filler interfaces (PU/THDBT) do not present a significant barrier to heat conduction. The interfacial thermal resistance in PU/THDBT is on the order of ~$10^{-9}$ m$^2$ K W$^{-1}$, as estimated by MD simulations—one to three orders of magnitude lower than that in polymer/inorganic filler hybrids.[30,60]

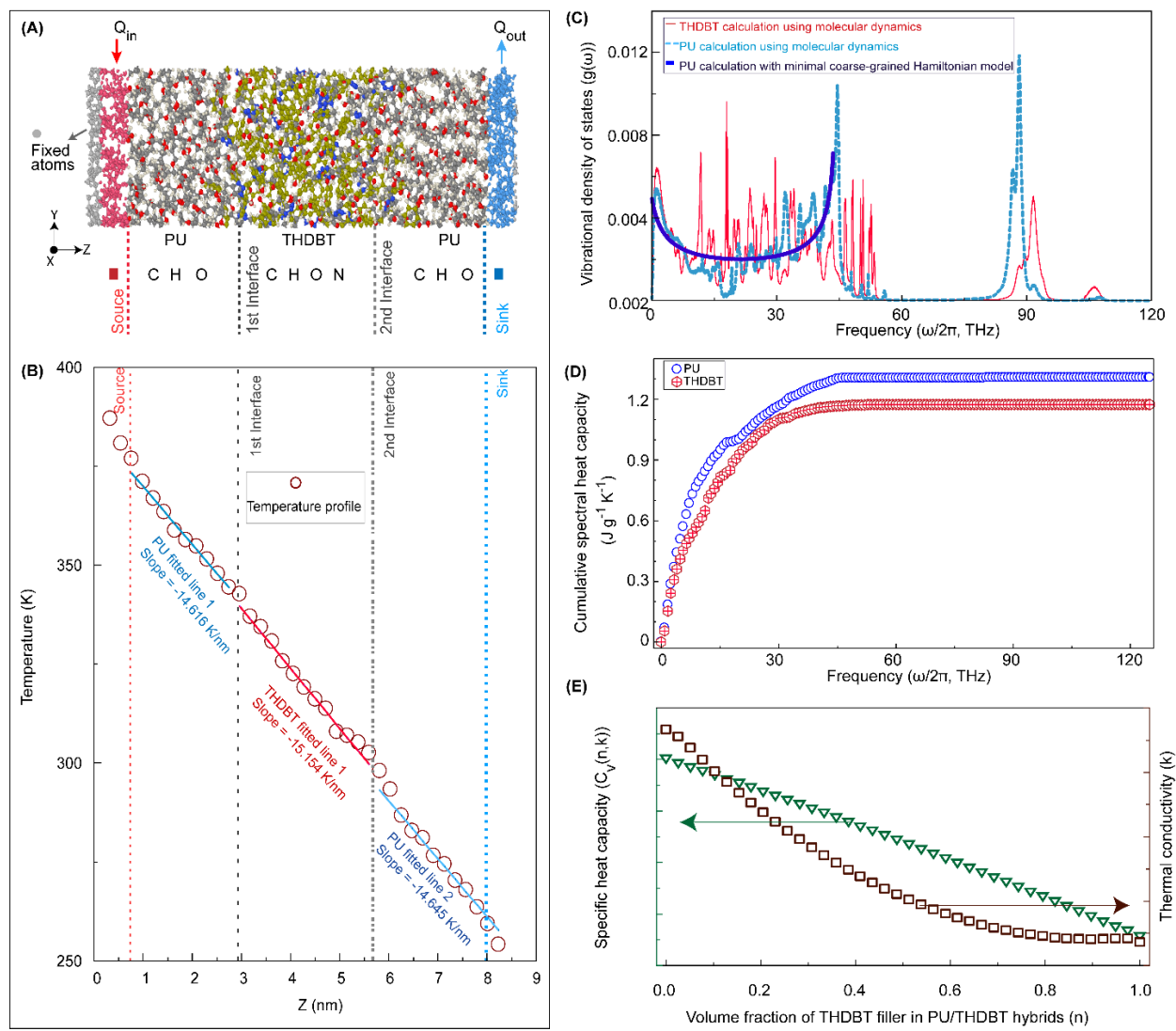

**Figure 3.** Theoretical and molecular dynamics studies conducted to reveal thermal transport mechanisms in nonporous polymer/organic filler hybrids. **(A)** Visual molecular dynamics (VMD) snapshot of PU/THDBT hybrid and the PU/THDBT interface. The volume fraction of THDBT is not relevant in this figure, as it corresponds to a single interface. **(B)** Temperature profile of PU/THDBT hybrid with fitted lines for PU and THDBT sections and labeled interfaces. **(C)** Vibrational density of states of PU and THDBT. **(D)** Molecular dynamic simulated relationships between cumulative spectral heat capacity and vibrational frequency of PU and THDBT. **(E)**

Calculated specific heat capacity and thermal conductivity as a function of the volume fraction of THDBT in PU/THDBT hybrids based on a minimal coarse-grained Hamiltonian model.

To better understand, at a theoretical level, why THDBT with rigid bonds exhibits a low specific heat capacity at the molecular scale, we analyze their vibrational density of states (VDOS) by MD simulations.[55,56] The specific heat capacity of THDBT is determined from their VDOS (**Figure 3C**). To obtain the VDOS, atomic velocities were recorded in the microcanonical (NVE) ensemble with a sampling interval of 4 fs over a total simulation time of 500 ps at 298 K and 1 atm. (**Figure S7** and Section S3 in the Supplementary Materials). The total vibrational contributions to the specific heat capacity of pure PU and pure THDBT were found to be 1.325 kJ kg$^{-1}$ K$^{-1}$ and 1.173 kJ kg$^{-1}$ K$^{-1}$, respectively. Despite its higher density (**Figures 2C and S7H**), THDBT exhibits a lower specific heat capacity than PU, consistent with experimental observations (**Figures 3D and 2B**). To investigate this difference, the spectral specific heat capacity ($\frac{\delta C_v}{\delta \omega}$) (**Figures S7I and S7J**) and the cumulative heat capacity (**Figure 3D**) are computed from the VDOS. **Figure 3D** reveals that vibrational modes with frequencies above 45 THz contribute negligibly to the heat capacity at 298 K, establishing 45 THz as an equivalent cutoff frequency for analyzing heat capacity. From the VDOS spectra (**Figure 3C**), the lower-frequency bands of PU extend up to this cutoff (45 THz), with broader peaks that encompass a larger spectral area below the cutoff frequency. In contrast, in THDBT, the lower frequency bands extend beyond the cutoff to ~55 THz, but the peaks are narrower and sharper, resulting in a smaller spectral area below the cutoff frequency ($<$ 45 THz). Consequently, THDBT contains fewer vibrational modes in the region that directly contributes to the heat capacity, leading to its lower specific heat capacity. Thus, the presence of high frequency modes and sharper peaks in the VDOS of THDBT are the primary factors underlying its lower specific heat capacity relative to PU. As a result, increasing the volume fraction of THDBT in PU/THDBT hybrids results in a reduction in the overall specific heat capacity.

To elucidate the origins of the differing specific heat capacities of THDBT and PU (**Figure 3D**), we investigated the respective roles of chemical structures in determining their vibrational contributions (**Figure S7K and S7L**). Decomposition of the phonon contributions by chemical group revealed that these high-frequency modes in THDBT originate from the benzene rings in its carbon backbone (**Figure S7L**). As a result, low frequency modes are comparatively less populated in THDBT than in PU. However, since high frequency phonons contribute little to the heat capacity, the overall specific heat capacity of THDBT is reduced (**Figure 3D**). In summary, the presence of benzene rings in THDBT shifts the phonon distribution toward higher frequencies, thereby lowering its heat capacity. Consequently, increasing the THDBT volume fractions in PU/THDBT hybrids lead to a reduction in the composite specific heat capacity.

To understand the origin of the broader peaks in PU, which enhance the heat capacity in the $0 - 45$ THz range (**Figure 3C**), we examined the role of chain structures in their vibrational contributions (**Figure S7K**). The aliphatic chains in PU (soft segments, **Figure 1B**) are flexible, allowing multiple conformations that spread vibrational modes over a wider and more continuous frequency range and broaden the VDOS peaks (**Figure S7K**). Thus, the aliphatic chains in PU generate broadened VDOS peaks owing to their chain flexibility and structural disorder. (**Figure S7K**). In contrast, benzene rings, with their rigid structures (**Figure 1D**), produce sharper and more discrete VDOS peaks (**Figure 7L**). This contrast highlights how chain rigidity governs vibrational features and, in turn, the heat capacity in the low frequency regime. Further details of MD simulations are given in section S3 in the supplementary information.

In parallel with the MD simulations, a minimal coarse-grained Hamiltonian model is developed to study how the THDBT filler concentration influences thermal transport in the PU/THDBT hybrids. The model reduces polymer thermal transport with randomly distributed fillers to a one-dimensional single-band phonon model analyzed using the coherent-potential approximation (CPA). Unlike the conventional virtual crystal approximation, the CPA accounts for multiple scattering and coherent effects, enabling direct calculation of physical observables in the filled polymer with higher fidelity, such as heat capacity and thermal conductivity. The results shown in **Figure 3E** closely reproduce the experimental trends (**Figures 2B and 2D**). Experimentally, both the specific heat capacity and thermal conductivity of PU/THDBT hybrids decrease with increasing filler volume fraction. Notably, despite the model's simplicity and limited parameters, it provides an excellent fit to the polymer VDOS (**Figure 3C**), supporting the validity of this minimal framework. Further details of the minimal coarse-grained Hamiltonian model are given in section S4 in the Supplementary Materials.

To quantitatively determine flame-retardant properties in PU/THDBT hybrids, three key flammability parameters are measured—heat release rate, heat release capacity, and fire growth capacity—using microscale combustion calorimetry (**Figure 4** and Section S2 in the Supplementary Materials).[61,62] Both pure PU and PU/THDBT hybrids (23.3 vol%) exhibited comparable onset decomposition temperatures of ~275 °C (**Figure 4A**). Pure PU films display two distinct thermal decomposition peaks,[63] which correspond to the degradation of different polymer segments (**Figure 4A**). The first peak is attributed to the decomposition of the hard segments, while the second peak is associated with the degradation of the soft segments in the PU chains.[64] This multi-stage decomposition behavior is consistent with previous studies.[64,65] The most compelling evidence of enhanced combustion resistance in the PU/THDBT hybrids is the significant reduction in the maximum heat release rate (**Figure 4A**).[66-68] Upon adding THDBT into the PU matrix, the maximum specific heat release rate decreased from ~497.00 W g$^{-1}$ in pure PU films to ~223.00 W g$^{-1}$ in the PU/THDBT hybrid (23.3 vol%), representing a reduction of about 55.13% (**Figure 4A**). At the same THDBT loading, the heat release capacity reduced from 497.00 ± 11.80 J g$^{-1}$ K$^{-1}$ to 223.00 ± 7.05 J g$^{-1}$ K$^{-1}$ (**Figure 4B**), while the fire growth capacity reduced from 333.00 ± 3.83 J g$^{-1}$ K$^{-1}$ to 242.00 ± 0.87 J g$^{-1}$ K$^{-1}$ (**Figure 4C**). These results suggest a notable improvement in the flame retardancy of the PU films upon addition of THDBT fillers. This enhancement is attributed to THDBT's ability to promote char formation and improve thermal stability, due to the synergistic effects of its dihydroxy-deoxybenzoin and triazole functional groups.[62] These groups facilitate the formation of a protective char barrier on the polymer surface, which limits heat transfer and reduces heat release during combustion.[62]

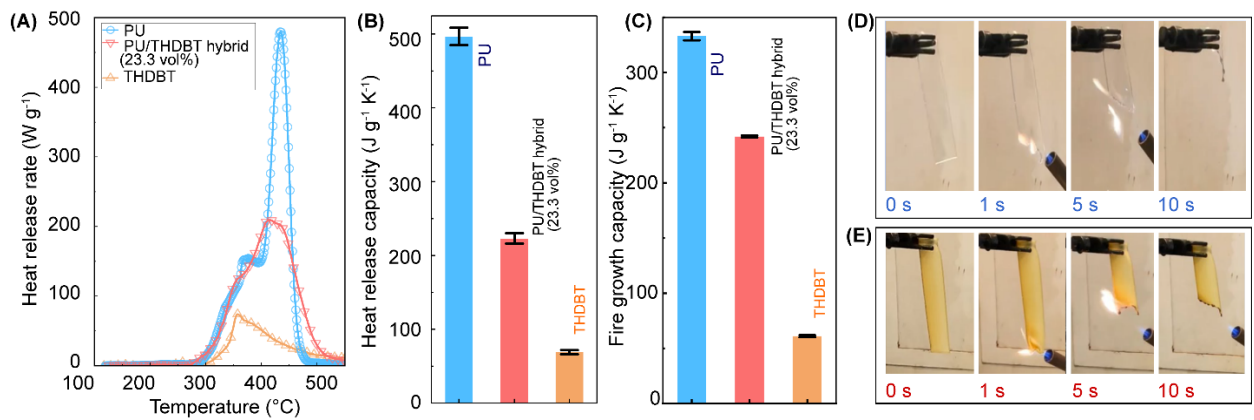

**Figure 4.** Experimental investigation of flame-retardant properties in polymers, organic fillers, and nonporous polymer/organic filler hybrids. **(A)** Experimental heat release rates for pure PU films, flame-retardant THDBT fillers, and PU/THDBT hybrid films (23.3 vol%). **(B)** Heat release capacity for pure PU films, flame-retardant THDBT fillers, and PU/THDBT hybrid films (23.3 vol%). **(C)** Fire growth capacity for pure PU films, flame-retardant THDBT fillers, and PU/THDBT hybrid films (23.3 vol%). **(D)** Vertical burning tests of pure PU films. **(E)** Vertical burning tests of PU/THDBT hybrid films (23.3 vol%). The sample films in Figures 5D and 5E were cut to dimensions of 10 cm × 2 cm × 0.01 cm (length × width × thickness) for the evaluation. A propane flame was applied under standard atmospheric conditions at 298 K. Each sample was exposed to the flame starting at 0 seconds, and the flame was removed after 5 seconds. The burning duration and extent were recorded using a stopwatch and camera to assess flammability performance.

To further evaluate the flammability, a vertical burning test is conducted on a pure PU film and a PU/THDBT hybrid (23.3 vol%), both having the same dimensions (**Figures 4D and 4E**).[62] The test is performed under standard atmospheric conditions (298 K, 1 atm) using a propane flame. Each specimen is exposed to the flame starting at 0 seconds, and the flame is removed after 5 seconds. The pure PU film ignited quickly and continued to burn intensely, eventually degrading significantly by 10 seconds (**Figure 4D**). In contrast, the PU/THDBT hybrid (23.3 vol%) exhibits improved flame resistance, showing limited ignition and charring while maintaining its structural integrity throughout the test (**Figure 4E**).

As discussed above, uniform THDBT filler dispersion and heterogeneous domain engineering within PU polymer play a critical role in understanding the thermal transport and flame retardancy properties of nonporous polymer/organic filler hybrids. To examine whether THDBT are uniformly dispersed within PU, carbon-13 solid-state nuclear magnetic resonance ($^{13}$C NMR) is employed (**Figure 5A**). These $^{13}$C NMR spectra are recorded under cross-polarization magic-angle spinning (CP/MAS) conditions. The hard segments of PU are derived from hexamethylene diisocyanate (HMDI, **Figure 5A**).[69] The soft segment of PU is polytetramethylene oxide (PTMO), which gives rise to the two tallest peaks in the NMR spectrum (**Figure 5A**).[70] The NMR spectrum of THDBT shows broad peaks, indicating that THDBT is largely amorphous.[62,71,72] The amorphous structures suggested by the NMR results are also consistent with the X-ray scattering data, which will be discussed below. The spectrum of the PU/THDBT hybrid (23.3 vol%) appears to be a simple mathematical sum of the PU and THDBT spectra, suggesting minimal interaction between the two components in the hybrid system. The $T_1$ relaxation time (~1.83 s) of THDBT in the PU/THDBT hybrid (23.3 vol%) is significantly shorter than that of neat THDBT (~2.8 s) and approaches the $T_1$ values observed for PU, indicating reduced domain sizes of THDBT within the hybrid and suggesting enhanced nanoscale dispersion (**Table S1** in the Supplementary Materials). This THDBT nanoscale dispersion is consistent with X-ray scattering results, which show a THDBT domain size of ~21 nm (**Figure S1**).

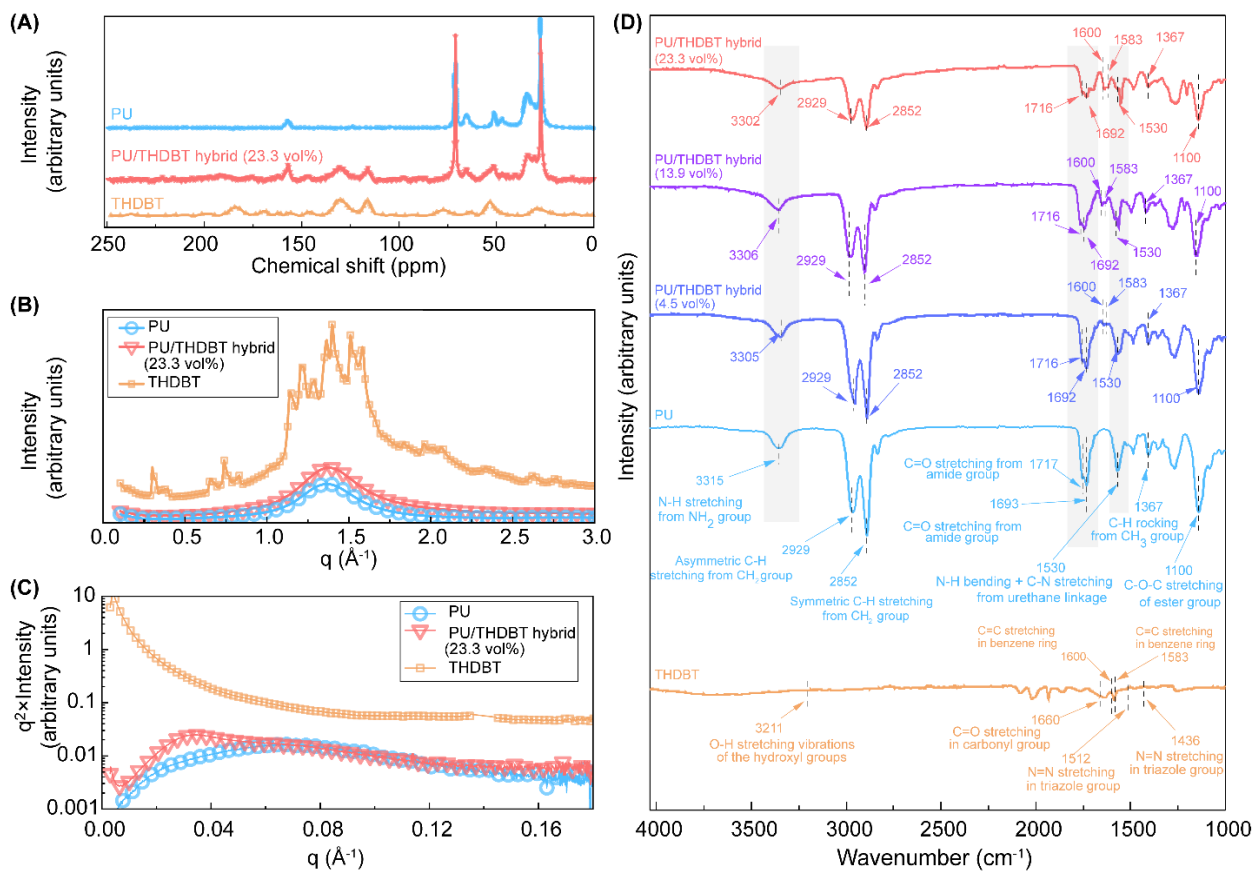

**Figure 5**. Experimental investigation of molecular structures and vibrational characteristics in polymers, organic fillers, and nonporous polymer/organic filler hybrids. **(A)** $^{13}$C CP/MAS NMR spectra of pure PU film, THDBT fillers, and PU/THDBT hybrids (23.3 vol%). **(B)** One-dimensional WAXS of pure PU films, compressed pellets of THDBT fillers, and PU/THDBT hybrids (23.3 vol%). **(C)** Kratky plot of SAXS of pure PU films, compressed pellets of THDBT fillers, and PU/THDBT hybrids (23.3 vol%). **(D)** ATR-FTIR spectra of the pure PU film; compressed pellets of THDBT fillers and potassium bromide (KBr); PU/THDBT hybrids (4.5 vol%), PU/THDBT hybrids (13.9 vol%), and PU/THDBT hybrids (23.3 vol%).

To better understand relationships between structures and thermal properties, we use synchrotron wide-angle X-ray scattering technique to investigate the lattice structures of PU films, compressed pellets of THDBT fillers, and PU/THDBT hybrids (WAXS; **Figures 5B, S1A, S1B and S1C**). **Figure 5B** presents one-dimensional WAXS intensity profiles as a function of the scattering vector ($q$) for PU, compressed pellets of THDBT fillers, and PU/THDBT hybrids (23.3 vol%). **Figure 5B** shows a broad peak between 1.0 Å$^{-1}$ and 2.0 Å$^{-1}$ for the compressed pellets of THDBT fillers, confirming their amorphous nature, which likely arises from bulky side groups that hinder crystallization. These amorphous structural features are consistent with the $^{13}$C CP/MAS NMR results discussed earlier.

To better understand relationships between domain size and thermal transport properties, we use small-angle X-ray scattering to investigate the domain size within PU/THDBT hybrids and the size of the hard segments in PU films, (SAXS, **Figures 5C and S1D–S1F**). **Figure 5C** displays the corresponding one-dimensional SAXS profiles as a function of the scattering vector ($q$) for PU, the compressed pellets of THDBT fillers, and the PU/THDBT hybrid (23.3 vol%). To better highlight structural changes, Kratky plots were generated for PU, the compressed pellets of THDBT fillers, and the PU/THDBT hybrid (23.3 vol%). For pure PU, a scattering peak at q ≈ 0.07 Å$^{-1}$ corresponds to a characteristic PU's hard segment size of ~9 nm. Upon incorporation of THDBT fillers, the peak at q ≈ 0.07 Å$^{-1}$—corresponding to the PU hard segment size of ~9 nm—remains present. In addition, a new peak emerges at q ≈ 0.03 Å$^{-1}$ in the PU/THDBT hybrid (23.3 $vol$%), corresponding to a domain size of ~ 21 nm (**Figure 5C**). This suggests that THDBT forms aggregated domains within the PU, most likely driven by intermolecular hydrogen bonding among phenolic hydroxyl groups, as well as between the phenolic –OH groups and the triazole moieties of THDBT.

To probe intermolecular interactions between THDBT and PU, we employed attenuated total reflectance Fourier transform infrared spectroscopy to examine hydrogen bonding within the PU/THDBT hybrids, (ATR-FTIR, **Figure 5D**).[43,73] In the ATR-FTIR spectra of pure THDBT (**Figure S8**), a broad peak at 3211 cm$^{-1}$ is observed, corresponding to the O–H stretching vibrations of the hydroxyl groups.[74,75] The broadness of this peak indicates strong and extensive hydrogen bonding among the hydroxyl groups within THDBT molecules.[76] This suggests the formation of molecular aggregates through self-association.[74,75] A peak at 1660 cm$^{-1}$ corresponds to the C=O stretching vibrations of the carbonyl groups in pure THDBT.[77] A peak at 1436 cm$^{-1}$ is assigned to the N=N stretching vibrations in the triazole groups of THDBT.[78] Peaks at 1600 cm$^{-1}$ and 1583 cm$^{-1}$ corresponds to the C=C stretching vibrations of the benzene rings of THDBT (**Figure S8**).[79] A peak at 1512 cm$^{-1}$ corresponds to the N=N stretching vibrations within the triazole groups in pure THDBT.[80] In the ATR-FTIR spectra of pure PU films (**Figure 5D**), a peak at 3315 cm$^{-1}$ appears, corresponding to the N–H stretching vibrations of amine groups located in the hard segments of PU.[81] The peak centered at 1717 cm$^{-1}$ corresponds to the C=O stretching vibrations of free carbonyl groups in PU, whereas the peak at 1693 cm$^{-1}$ corresponds to the C=O stretching vibrations of hydrogen-bonded carbonyl groups, both within the PU hard segments.[82,83] The peak at 1530 cm$^{-1}$ arises from the amide II band associated with urethane linkages, comprising N–H bending and C–N stretching vibrations.[84] Peaks centered at 2929 cm$^{-1}$ and 2852 cm$^{-1}$ are attributed to the asymmetric and symmetric stretching vibrations of CH$_2$ groups, respectively.[85,86] In the ATR-FTIR spectra of PU/THDBT hybrids (**Figure 5D**), the stretching vibration peaks of the amine group (N–H) in PU shift depending on the THDBT volume fractions, compared to those in pure PU. Specifically, the N–H stretching peak shifts from 3315 cm$^{-1}$ to 3302 cm$^{-1}$ for PU/THDBT hybrids (23.3 vol%). These shifts suggest the presence of hydrogen bonding interactions between PU and THDBT.[82] We acknowledge that the PU/THDBT hybrids exhibited some degree of non-uniformity during ATR-FTIR analysis, which necessitated multiple measurements to ensure repeatability and reliability.

To summarize the ATR-FTIR results, we observed the presence of intermolecular hydrogen bonding among PU hard segments, THDBT aggregates, and between PU hard segment domains and THDBT aggregate domains. These hydrogen-bonded intermolecular interactions are expected to facilitate the formation of heterogeneous domain boundaries and promote the uniform dispersion of THDBT within the PU matrix, enabling the investigation of interfacial effects on thermal transport and flame retardancy in nonporous polymer/organic hybrids.

**Conclusion**

Using PU/THDBT hybrids as a model system, this work demonstrates for the first time that the reduction in thermal conductivity in nonporous polymer/organic hybrids originates primarily from the suppression of specific heat capacity, enabled by controlling atomic vibrations through engineered chemical compositions and structural design. Compressed pellets of THDBT fillers exhibit a ~55% lower specific heat capacity than PU films, attributed to rigid bonds in benzene-rings and triazole-rings of THDBT that restrict vibrational freedom of atoms in these rings. This intrinsically low heat capacity leads to a 17% reduction in thermal conductivity of PU/THDBT hybrids (23.3 vol%) compared to pure PU. Integrated experiments and MD simulations reveal how atomic vibrations shape the vibrational density of states, which in turn determines the specific heat capacity. The atoms connected by rigid bonds (e.g., double bonds) in benzene-rings and triazole-rings of THDBT generate narrow and discrete vibrational modes that reduce the low-frequency spectral area (0–45 THz), resulting in lower heat capacity. In contrast, atoms connected by flexible bonds (e.g., single bonds) in flexible aliphatic chains in PU generate broad and continuous vibrational modes that enhance the low-frequency spectral area (0–45 THz), resulting in higher heat capacity. A coarse-grained minimal Hamiltonian model based on the coherent potential approximation semi-quantitatively reproduces the experimentally observed filler concentration dependence of both heat capacity and thermal conductivity and successfully captures MD-simulated VDOS trends. MD simulations estimate the PU/THDBT interfacial thermal resistance to be on the order of $\sim 10^{-9}$ m$^2$ K W$^{-1}$. This interfacial thermal resistance between PU and THDBT is negligible and does not significantly hinder heat conduction in polymer/organic filler hybrid. The results in this work contrast with the conventional belief that interfacial thermal resistance hinders heat conduction in polymer/inorganic filler hybrids.

Furthermore, this study establishes for the first time a design framework for nonporous polymer/organic filler hybrids that concurrently reduce thermal conductivity and enhance flame retardancy. Hydrogen-bond interactions between PU segments and THDBT create heterogeneous domains that promote uniform dispersion and facilitate char formation during combustion. PU/THDBT hybrids with 23.3 vol% THDBT reduce fire growth capacity by ~27%. Unlike halogenated flame retardants, this eco-friendly strategy leverages fillers with intrinsic char-forming capability.

This work establishes a design framework for nonporous polymer/organic filler hybrids with tunable thermal transport and introduces a new design principle—suppressing filler-specific heat capacity—to achieve intrinsically low thermal conductivity through molecular-level control of chemical compositions, structures, and atomic vibrations. Building on these results, it advances the fundamental understanding of thermal transport in nonporous polymer/organic filler hybrids and guides the design of next-generation materials with tunable thermal and flame-retardant properties for energy-efficient and safe applications across insulation, aerospace, healthcare, and electronics.

**Conflict of Interest**

The authors declare that they have no conflict of interests.

## Materials

All reagents and solvents were purchased from the commercial supplier Sigma-Aldrich and Fisher Scientific. All chemicals were obtained commercially and used without further purification. Polyurethane (PU) has a molecular weight of ~177,900 g $mol^{-1}$. PU was purchased from Sigma-Aldrich. N,N-Dimethylformamide (DMF) was purchased from Sigma-Aldrich. Tetrahydrofuran (THF) was purchased from Fisher Scientific. THDBT was synthesized as described in the publication.[62]

Preparation of PU/THDBT hybrids at THDBT volume fractions of 4.5, 13.9, and 23.3 vol%. PU/THDBT hybrid films with THDBT volume fractions of 4.5, 13.9, and 23.3 vol% are prepared using a solution blending method. As a representative example, the preparation of a PU/THDBT hybrid film containing 23.3 vol% THDBT is described below. THDBT (0.067 g) and PU (0.200 g) were placed in a 20 mL glass vial, followed by the addition of N,N-dimethylformamide (DMF, 2.670 g) as the solvent. The mixture was heated to ~75 °C in an oil bath and magnetically stirred for 6 h until the PU was completely dissolved, yielding a homogeneous PU/THDBT solution. The solution was then cast into a silicone mold (100 mm × 20 mm) and dried in an oven at 75 °C for 24 h. Subsequently, the film was further dried under vacuum at 120 °C for 12 h. After drying, a flexible hybrid film with a thickness of ~100 μm was obtained by peeling it from the mold. It is noteworthy that the dispersion of THDBT within the PU matrix deteriorated significantly when the THDBT content exceeded 23.3 vol%. Therefore, 23.3 vol% is selected as the maximum filler volume fraction in this study. For clarity, these PU/THDBT hybrids are denoted as PU/THDBT (4.5 vol%), PU/THDBT (13.9 vol%), and PU/THDBT (23.3 vol%).

Details on the preparation of pure PU films and compressed pellets of THDBT fillers are provided in Section S1 in the Supplementary Materials.

## Author Contributions

H.W., M.C., and Y.Z. contributed equally to this work. Y.X. conceived the idea for the paper. H.W., M.C., Y.Z., K.M., W.H., S.W., and R.L. carried out experimental measurements. K.M., T.E., and Y.X. performed the polymer/filler design and synthesis. H.W., M.C., Y.Z., W. Y, and A.C. analyzed the measured thermal transport properties. H.W., M.C., Y.Z., W. Y, and W. H analyzed the measured structures. Z.B. and J.L. developed molecular dynamics simulation models and analyzed simulated thermal transport properties. M.C. constructed the theoretical model and performed the quantum transport calculations. S.W. and R.L. carried out synchrotron X-ray scattering experiments and data analysis. H.W., Z.B., J.L., M.C., and Y.X. wrote the original draft of the paper. All authors discussed the results and provided feedback on the manuscript.

## Acknowledgments

Funding: This work was funded by the Faculty Startup Fund support from the University of Massachusetts Amherst awarded to Y.X., Federal Aviation Administration (award number FAA 17-G-012) to Y.X., Elaine Marieb Center for Nursing and Engineering Innovation at the University of Massachusetts Amherst (award number 190427) awarded to Y.X., the National Science Foundation (award number 2312559) awarded to Y.X. the National Science Foundation (award number CBET-1943813) awarded to J.L. This research used resources of the National Synchrotron Light Source II, a U.S. Department of Energy (DOE) Office of Science User Facility operated for

# Supplementary Materials for

# Engineering Nonporous Polymer Hybrids with Suppressed Heat Conduction and Superior Flame Retardancy via Molecular and Filler Design

Henry Worden,[1#] Mihir Chandra,[2#] Yijie Zhou,[1#] Zarif Ahmad Razin Bhuiyan,[3] Mouyang Cheng,[4] Krishnamurthy Munusamy,[5] Weiguo Hu,[5] Weibo Yan,[1] Siyu Wu,[6] Ruipeng Li,[6] Anna Chatterji,[1] Todd Emrick,[5] Jun Liu,[3] Yanfei Xu[1,2*]

7. Department of Mechanical and Industrial Engineering, University of Massachusetts, Amherst, Massachusetts, 01003, United States of America
8. Department of Material Science and Engineering, University of Massachusetts, Amherst, Massachusetts, 01003, United States of America
9. Department of Mechanical and Aerospace Engineering, North Carolina State University, Raleigh, NC 27695, USA.
10. Department of Materials Science and Engineering, Massachusetts Institute of Technology, Cambridge, MA 02139, USA.
11. Department of Polymer Science and Engineering, University of Massachusetts, Amherst, Massachusetts, 01003, United States of America
12. Brookhaven National Laboratory, Upton, New York, 11973, United States of America

#Contribute equally

*Corresponding author yanfeixu@umass.edu

**Supplementary Materials**

This file includes:

- **Section S1.** Sample preparation
- **Section S2.** Experimental characterization and error analysis
- **Section S3.** Molecular dynamics simulations of thermal transport in polymer/filler hybrids
- **Section S4.** Effective Hamiltonian model for polymer thermal transport with fillers

**Figures S1–S10**
**Table S1**
**References**

## Section S1. Sample preparation, experimental characterization, and error analysis.

### Section S1.1 Sample preparations.

**Preparation of PU solution and pure PU thin films.**

A solution containing 7 $wt\%$ of PU was prepared using DMF as solvent. PU (0.2 $g$) was added into DMF (2.67 $g$) in a 20 $mL$ glass vial. The temperature of the mixture was then raised to ~ 75 °$C$ using an oil bath and stirred for 6 hours until the PU was completely dissolved. The solution was then poured into a silicone mold (100 $mm$ × 20 $mm$) and dried in an oven at 75 °$C$ for 24 hours. The sample was further dried under vacuum at 120 °$C$ for 2 hours to fully remove the DMF. The final PU film with thickness of ~80 μm was peeled off the silicone mold after the drying process completed.

**Preparation of tetrahydroxy deoxybenzoins triazole (THDBT).**

The detailed synthesis of THDBT has been described in the recent publication.[1] Briefly, the synthesis involves a two-step process:

(1) Preparation of tetramethoxy deoxybenzoins triazole.

1,4-Bis(azidomethyl)benzene (2.00 $g$, 10.6 $mmol$) and alkynyl desoxyanisoin (7.80 $g$, 26.6 $mmol$) were reacted in tetrahydrofuran (THF, 20 $mL$) under nitrogen. An aqueous solution of sodium ascorbate and CuSO$_4$·5H$_2$O was added. The mixture was heated at 30 °$C$ for 12 hours, then extracted with dichloromethane. The organic layer was dried over magnesium sulfate, filtered, and concentrated. The crude product was purified to yield tetramethoxy deoxybenzoins triazole as a yellow powder and was confirmed by $^1$H and $^{13}$C NMR spectroscopy.

(2) Conversion to THDBT.

Tetramethoxy deoxybenzoins triazole (5.00 $g$, 6.43 $mmol$) was treated with hydroiodic acid (14 $mL$) in glacial acetic acid (*10 mL*) at 140 °C for 7 hours. After cooling, the mixture was poured into water to precipitate THDBT, which was then filtered and dried. Finally, the compound was confirmed by $^1$H and $^{13}$C NMR spectroscopy.[1]

**Preparation of samples for thermal diffusivity measurements.**

(1) Preparation of compressed pellets of THDBT fillers for thermal diffusivity measurements.

To measure the cross-plane thermal diffusivity of THDBT fillers, compressed pellets of THDBT fillers are prepared using the following procedure. The THDBT powder is weighed out (~0.7 $g$) as per experimental design and is poured into a customized mold with a diameter of 12.7 mm. A pressure of 58 MPa is applied to the mold using a hydraulic press for 3 minutes to form the pellet specimen (thickness ~0.7 $mm$). The compressed pellets of THDBT fillers are then sprayed with graphite (DGF 123, Miracle Power Products) prior to measuring cross-plane thermal diffusivities.

(2) Preparation of pure PU films for thermal diffusivity measurements.

PU/THDBT hybrid films with THDBT volume fractions of 4.5, 13.9, and 23.3 $vol\%$ were prepared via a solution blending method. As a representative example, the preparation of a film containing 23.3 $vol\%$ THDBT is described. THDBT (0.067 $g$) and PU (0.200 $g$) were placed in a 20 $mL$ glass vial, followed by the addition of N,N-dimethylformamide (DMF, 2.670 $g$) as the solvent. The mixture was heated to ~75 °$C$ in an oil bath and magnetically stirred for 6 h until the PU was completely dissolved, forming a homogeneous PU/THDBT solution. The solution was then cast into a silicone mold (100 $mm$ × 20 $mm$) and dried in an oven at 75 °$C$ for 24 hours, followed by vacuum drying at 120 °$C$ for 12 hours. After drying, a flexible hybrid film (~100 μm thick) was obtained by peeling it from the mold.

(3) Preparation of pure PU films for thermal diffusivity measurements.

To measure the cross-plane thermal diffusivity of pure PU films, samples are prepared as follows. PU (0.2 $g$) was dissolved in N,N-dimethylformamide (DMF, 2.67 $g$) in a 20 $mL$ glass vial. The mixture was heated to ~75 °$C$ in an oil bath and stirred for 6 hours until the PU was completely dissolved. The resulting solution was cast into a silicone mold (100 $mm$ × 20 $mm$) and dried in an oven at 75 °$C$ for 24 hours, followed by vacuum drying at 120 °$C$ for 2 hours to remove residual solvent. The obtained PU film (~80 $\mu m$ thickness) was removed from the mold and cut into circular specimens (12.7 $mm$ in diameter) using a hollow punch.

**Preparation of samples for X-ray scattering measurements.**

(1) Preparation of PU thin films for X-ray scattering measurements.

Samples were prepared using the same procedure as those used for the thermal diffusivity measurements.

(2) Preparation of PU/THDBT hybrids for X-ray scattering measurements.

Samples were prepared using the same procedure as those used for the thermal diffusivity measurements.

(3) Preparation of compressed pellets of THDBT fillers for X-ray scattering measurements.

Samples were prepared using the same procedure as those used for the thermal diffusivity measurements.

**Preparation of samples for Attenuated total reflectance fourier transform infrared (ATR-FTIR) spectroscopic analysis**

(1) Preparation of PU thin films for ATR-FTIR spectroscopic analysis (Figure 5D).

Samples were prepared using the same procedure as those used for the thermal diffusivity measurements.

(2) Preparation of PU/THDBT hybrids for ATR-FTIR spectroscopic analysis (Figure 5D).

Samples were prepared using the same procedure as those used for the thermal diffusivity measurements.

(3) Preparation of compressed pellets of THDBT fillers for ATR-FTIR spectroscopic analysis (Figure 5D).

Samples were prepared using the same procedure as those used for the thermal diffusivity measurements.

(4) Preparation of THDBT fillers for ATR-FTIR spectroscopic analysis (Figure S8).

Approximately 24 $mg$ of THDBT powder was mixed with 71 $mg$ of potassium bromide (KBr) and finely ground to ensure homogeneity. The mixture was then loaded into a custom mold (12.7 mm in diameter) and pressed at 46 $MPa$ for 1 minute using a hydraulic press to form pellet specimens.

## Section S2. Experimental characterization and error analysis

### Section S2.1 Experimental characterizations and instrumental measurement details

**Thermal diffusivity measurements.**

(1) Cross-plane thermal diffusivities of pure PU films and PU/THDBT hybrid (films) measured using a laser flash apparatus (LFA 467 HyperFlash, NETZSCH). The instrument is operated with a lamp voltage of 150 $V$ and a pulse width of 50 $\mu s$. Prior to measurement, both sides of each sample (prepared as described in Section S1.1) were coated with graphite spray (DGF 123, Miracle Power Products) and allowed to dry at room temperature for 5 $minutes$.[2] Prior to measurement, all samples were equilibrated at 298 $K$. For PU films and PU/THDBT hybrid films, the transparent model in the NETZSCH analysis software is applied to evaluate the laser flash data and determine the thermal diffusivity.[2]

(2) Cross-plane thermal diffusivities of compressed pellets of THDBT fillers

The cross-plane thermal diffusivities of the compressed pellets of THDBT fillers are measured using a laser flash apparatus (LFA 467 HyperFlash, NETZSCH). The instrument is operated with a lamp voltage of 250 $V$ and a pulse width of 600 $\mu s$. Prior to measurement, both sides of each sample (prepared as described in Section S1.1) were coated with graphite spray (DGF 123, Miracle Power Products) and allowed to dry at room temperature for 5 min.[2] Prior to measurement, all samples were equilibrated at 298 $K$. For compressed pellets of THDBT fillers, the penetration model was used for data analysis and thermal diffusivity extraction.

**Specific heat capacity measurements.**

Specific heat capacities are measured using differential scanning calorimetry (DSC 2500, TA Instruments). Measurements are conducted from 273 $to$ 323 $K$ at a heating rate of 10 $°C\ minute^{-1}$. Each DSC run consisted of four heating–cooling cycles; the first cycle is used to remove the sample's thermal history, and the subsequent three cycles are used to determine the specific heat capacity. Approximately $6 - 9\ mg$ of samples is used per measurement. Nitrogen is employed as the purge gas at a flow rate of 300 $mL\ minute^{-1}$. Film samples prepared in Section S1 are cut into circular specimens (6.7 $mm$ in diameter) using a hollow punch.

**Density measurements by Archimedes' principle.**

The densities of pure PU films, PU/THDBT hybrids, and compressed pellets of THDBT fillers are determined using Archimedes' principle. *n*-hexane is used as the immersion medium. Approximately 0.5 $g$ of each sample was placed in a graduated cylinder, followed by the addition of 5 mL of *n*-hexane. The mixture was allowed to settle for 10 $minutes$, and the increase in liquid volume corresponded to the sample volume. Each measurement was repeated three times to obtain an average density value for each sample.[2]

**$^{13}$C solid-state nuclear magnetic resonance (NMR) analysis.**

A saturation-recovery experiment was conducted to study the $^1$H T$_1$ relaxation of each constituent in the samples. Saturation was achieved by a train of sixteen $^1$H 90º pulses spaced by 1 ms delays, followed by a variable recovery period ranging between $50\ ms$ and $30\ s$, then followed by a $^1H$ 90º excitation, $^1H$-to-$^{13}C$ cross polarization (CP) and $^{13}C$ detection with $^1H$ decoupling. The experiments were performed under magic angle spinning (MAS) on a Bruker 600 MHz solid-state NMR spectrometer in a $4\ mm$ broadband-observe CP/MAS probe. A spinning speed of $9\ kHz$, contact time of $2\ ms$, a recycle delay of $5\ s$, and a decoupling field strength of $60\ kHz$ were used for NMR experiments. Chemical shift was calibrated by setting the unprotonated aromatic carbon signal of 1,4-di(t-butyl)benzene at 148.8 ppm.

**Synchrotron X-ray scattering measurement.**

Wide-angle and small-angle X-ray scattering (WAXS/SAXS) experiments were performed at the Complex Materials Scattering (CMS, 11-BM) beamline at the National Synchrotron Light Source II (NSLS-II), Brookhaven National Laboratory. Samples are mounted in a custom-designed static holder configured for transmission geometry. X-rays with an energy of $13.5\ keV$ ($\lambda \approx 0.918\ Å$) are used. Scattering patterns are collected with Pilatus 2M (SAXS) and Pilatus 800K (WAXS) detectors positioned at sample-to-detector distances of $\sim 5\ m$ and $\sim 0.26\ m$, respectively. The incident beam size is approximately $0.2\ mm\ \times\ 0.2\ mm$. Background subtraction was carried out using scattering data from the empty holder. Two-dimensional scattering images are azimuthally integrated using the SciAnalysis software package to obtain one-dimensional $I(q)$ versus $q$ profiles. The exposure time for each measurement is $30\ s$.

**Attenuated total reflectance Fourier transform infrared (ATR-FTIR) spectroscopy analysis.**

The ATR-FTIR spectroscopy is performed using a PerkinElmer Spectrum One FT-IR spectrometer. Spectra are recorded in the range of $4000 - 550\ cm^{-1}$ with a resolution of $2\ cm^{-1}$. To improve the signal-to-noise ratio, each sample is scanned ten times, and the averaged spectrum is used for analysis. All measurements are conducted at $298\ K$ under ambient conditions.

**Microscale combustion calorimeter analysis.**

The microscale combustion calorimeter analysis was performed according to the ASTM D7309-21, method A,[3] using an $80\ cm^3\ min^{-1}$ stream of N2(g) and a heating rate of $1\ °C\ s^{-1}$. In the microscale combustion calorimeter, the anaerobic thermal degradation products were combined with a $20\ cm^3\ min^{-1}$ stream of oxygen gas in a furnace at $900\ °C$. Key flammability parameters, including heat release capacity, fire growth capacity, and total heat release were calculated from experimental data.[4-6]

Heat release capacity quantifies the maximum heat release rate normalized by the mass loss rate during combustion and serves as a reliable indicator of a material's intrinsic flammability, largely dictated by its chemical structure and degradation behavior (Figure 4B).[1,7] Fire growth capacity provides a broader evaluation by incorporating both combustibility and ignitability, especially reflecting how rapidly a polymer contributes to fire growth following ignition (Figure 4C).[1,7] Together, heat release capacity and fire growth capacity offer complementary metrics that enable a comprehensive assessment of flammability in polymer materials.

### Section S2.2 Error analysis.

Population standard deviation and error propagation analyses of thermal diffusivity, specific heat capacity, density, and thermal conductivity were performed according to previous publications.[2]

## Section S3. Molecular dynamics simulations of thermal transport in polymer/filler hybrids.

### Section 3.1 Molecular dynamics simulations – methodology.

Atomistic molecular dynamics (MD) simulations were conducted to evaluate the interfacial thermal resistance and specific heat capacity of the PU/THDBT hybrids. All simulations were performed using LAMMPS[8] and structural visualization was carried out with OVITO.[9]

Three simulation systems were considered in this study: pure PU, pure THDBT, and the PU–THDBT interface as shown in Figure S7(A), (C)and (E)respectively. The total number of atoms in these systems was 5610, 4950, and 4986, respectively. The atomic configurations were generated under the assumption that each polymer was in a purely amorphous state. The polymer consistent force field (PCFF) was employed, with a time step of $0.5\ fs$. Each system was first equilibrated for 1 ns under the NPT ensemble at $298\ K$ and $1\ atm$. After NPT relaxation, the simulation box dimensions were $2.60\ \times\ 2.602\ \times\ 8.21\ nm$ for PU, $2.41\ \times\ 2.41\ \times\ 8.75\ nm$ for THDBT, and $2.67\ \times\ 2.67\ \times\ 7.45\ nm$ for the PU/THDBT interface.

### Section 3.2 Determination of interfacial thermal resistance and thermal conductivity

A highly localized section of the PU/THDBT matrix was extracted to model the interfacial region. Two PU/THDBT interfaces were constructed by placing a THDBT layer between two PU regions, as illustrated in **Figure S7 (E)**. The interfacial thermal resistance was evaluated using the non-equilibrium molecular dynamics (NEMD) approach. For this purpose, the simulation domain along the $z$-axis was partitioned into multiple regions. Total $0.22\ nm$ layer at each end of the simulation box was designated as a fixed region, where atomic forces and velocities were set to zero, thereby rendering the system aperiodic along the $z$-direction. Periodic boundary conditions were applied in the $x$ and $y$ directions. Adjacent to the left and right fixed regions, $5.5\ nm$ zones were defined as the heat source and heat sink, respectively, for imposing the thermal gradient. The system was first equilibrated under the NVT ensemble for 1 ns, after which it was switched to the NVE ensemble. In the NVE stage, heat was continuously supplied to the source region and removed from the sink region at a constant rate of $17.36\ nJ\ s^{-1}$. This heating–cooling process was maintained for 6 ns, until the average temperatures of the source and sink reached a steady state, as shown in **Figure S7 (G)**. The simulation box was then divided into 40 equally spaced slabs along the $z$-direction, and the temperature and density of each slab were computed and averaged over an additional 1 ns. The resulting spatial variations in density and temperature are presented in **Figure S7 (H)** and **Figure 3B** in the main text respectively, and are hereafter referred to as the density and temperature profiles. The positions of the interfaces were identified by visual inspection of these profiles, corresponding to abrupt changes in both density and temperature. The interfaces were located at $2.93\ nm$ and $5.68\ nm$ from the origin along the $z$-axis, as indicated by the dotted vertical lines in **Figure 3B** of the main text. To quantify the temperature discontinuity across the interfaces, linear fits were applied to the temperature profile within the three distinct regions (PU, THDBT, and PU), and the temperature drops were extracted from the intersections of these fitted lines with vertical lines of the interfaces, as illustrated in **Figure 3B** of the main text. The interfacial thermal resistance (ITR) was calculated using equation S3-Eq1 in this supplementary materials file.

$$ITR = \frac{\Delta T A}{\dot{Q}} \quad \text{(S3-Eq1)}$$

where $\Delta T$ is the temperature drops across the interface, A is the cross-sectional area of the xy-plane perpendicular to the heat-transfer direction, and $\dot{Q}$ is the heat rate supplied to the source or removed from the sink.

A similar procedure was employed to determine the thermal conductivity of the pure PU and pure THDBT systems. In this case, instead of measuring the interfacial temperature drop, the slope of the fitted linear segment in the temperature profile (data not shown) was obtained. The thermal conductivity was then calculated using Fourier's law of heat conduction as expressed in Equation S3-Eq2.

$$\kappa = -\frac{\dot{Q}}{A\frac{dT}{dz}} \quad \text{(S3-Eq2)}$$

Where, $\kappa$ is thermal conductivity, $\dot{Q}$ is heat rate applied to heat source or removed from the heat sink, $A$ is the xy plane cross-sectional area and $\frac{dT}{dz}$ is the slope of the temperature profile along heat transfer direction.

### Section 3.3 Determination of specific heat capacity.

The specific heat capacities of pure PU and pure THDBT were determined from their vibrational density of states (VDOS). To obtain the VDOS, atomic velocities were recorded under the NVE ensemble with a sampling interval of $4\ fs$ over a total simulation time of $500\ ps$. The VDOS was calculated by performing a Fourier transform of the velocity time series for each velocity components, followed by computing the mass-weighted average magnitude of the transformed velocities across all atoms. The computed VDOS spectra was normalized such that the total area under the curve equaled to unity. The mathematical formulation for the computation of VDOS is expressed in Equations S3-Eq3 through *S3-Eq5*.[10,11]

$$\hat{v}_p(\omega) = \int_{-\infty}^{\infty} v_p(t)\, e^{-i\omega t} dt \quad \text{(S3-Eq3)}$$

$$g(\omega) = \frac{A}{N_{atom}} \sum_{i=1}^{N_{atom}} m_i \sum_{p=1}^{3} [\hat{v}_p(\omega) \times \hat{v}_p^*(\omega)] \quad \text{(S3-Eq4)}$$

$$\int g(\omega)\, d\omega = 1 \quad \text{(S3-Eq5)}$$

Where, $g$ is the normalized magnitude of VDOS, $\omega$ is angular frequency, A is the normalization prefactor, $v_p\ (p = 1,2,3)$ is the velocity components of an atom $i$ ($i = 1,2,\ldots N_{atom}$), $m$ is the mass of the atom, $N_{atom}$ is the total number of atoms in the system and $t$ denotes time.

Finally, specific heat was calculated from Equation S3-Eq6.[11,12]

$$C_v(T) = \frac{3N_{atoms}k_b}{m_{system}} \int \left(\frac{\hbar\omega}{k_bT}\right)^2 \frac{\exp\left(\frac{\hbar\omega}{k_bT}\right)}{\left[\exp\left(\frac{\hbar\omega}{k_bT}\right)-1\right]^2} g(\omega)d\omega \quad \text{(S3-Eq6)}$$

Where, $C_v$ is the specific heat at constant volume, $k_b$ is the Boltzmann constant, $\hbar$ is reduced Planck's constant, $T$ is the temperature, $m_{system}$ is the total mass of the system and $N_{atom}$ is the total number of atoms in the system.

### Section 3.4 Results and discussion

Computed density of PU and THDBT after NPT relaxation at 298 $K$ temperature and 1 atm pressure are 0.94 $g\ cm^{-3}$ and 1.24 $g\ cm^{-3}$ Density of THDBT is higher than PU which follows the experimental trends where experimental values are 1.095 $g\ cm^{-3}$ and 1.15 $g\ cm^{-3}$, respectively.

Thermal conductivity of pure PU and THDBT from the simulation are 0.206 $W\ m^{-1}\ K^{-1}$ and 0.199 $W\ m^{-1}\ K^{-1}$, which also follows the experimental trends but does not match the experimental values.

Interfacial thermal resistances (R$_{ITR}$) at two interfaces are $0.5 \times 10^{-9}$ and $1.1 \times 10^{-9}$ $m^2\ K\ W^{-1}$ respectively which corresponds to 1.49 $K$ and 2.97 $K$ temperature drop at interface. Average R$_{ITR}$ is $0.80 \times 10^{-9}\ m^2\ K\ W^{-1}$. These values suggest that there is very low thermal resistance between PU and THDBT.

Although the simulation results follow the experimental trends, they do not exactly match the experimental values. Several factors may contribute to these discrepancies. First, in the MD simulations, PU was modeled as an amorphous phase, whereas in the experiments PU exhibits a polycrystalline structure. Second, there is a size effect: in MD simulations, only a highly localized (nanoscale) portion of the material is modeled, and it is well established that thermal properties are size-dependent at this scale. The combined influence of these factors leads to deviations from experimental values, even though the simulations can reliably predict the overall experimental trends.

### Section S4. Effective Hamiltonian model for polymer thermal transport with fillers.

### Section 4.1 Hamiltonian model setup for a polymer chain.

We consider the 1D polymer chain, with monomers and their on-site monomer energy $\varepsilon_A$ and inter-monomer hopping $t_A$, the pristine polymer's (PU) vibrational Hamiltonian can be written as

$$H_{PU} = \sum_j \varepsilon_A a_j^+ a_j - t_A \sum_j (a_j^+ a_{j+1} + \text{h.c.}) \quad \text{(S4-Eq1)}$$

where for simplicity, we neglect the hard/soft segments, but only treat PU as a uniform polymer, and the standard Bosonic commutation relation $[a_i, a_j^+] = \delta_{ij}$ applies. After Fourier transform $a_k = \frac{1}{\sqrt{N}} \sum_{j=1}^{N} e^{-ikaj} a_j$ (with inter-monomer spacing $a$), the pristine polymer's vibrational Hamiltonian can be diagonalized as

$$H_{PU} = \sum_k [\varepsilon_A - 2t_A \cos ka] a_k^+ a_k \quad \text{(S4-Eq2)}$$

Now, the fire-retardant THDBT blocks (termed as "B") are randomly distributed in the polymer chain with a filling fraction $c$. We treat such monomer-THDBT composite as binary alloy system, with both diagonal, on-site disorder as well as off-diagonal disorder from hopping. The composite Hamiltonian can be written as

$$H = \sum_j \varepsilon_j a_j^+ a_j - \sum_j t_{j,j+1}(a_j^+ a_{j+1} + \text{h.c.}) \quad \text{(S4-Eq3)}$$

where $\varepsilon_j \in \{\varepsilon_A, \varepsilon_B\}$, $t_{j,j+1} \in \{t_{AA}, t_{BB}, t_{AB}\}$. Here "A" denotes the original monomers with concentration 1-c, and "B" denotes the THDBT blocks with concentration c.

### Section 4.2 BEB-coherent potential approximation.

To proceed, we adopt the coherent-potential approximation (CPA) with an extension off-diagonal disorder, i.e. disorder from the hopping. This is the so-called Blackman-Esterling-Berk (BEB) formalism as a generalization of CPA.[13] The effective "coherent" Hamiltonian $H_0$, which serves as a background effective medium, can be written as

$$\begin{aligned} H_0 &= \sum_j \varepsilon_{\text{coh}}(\omega) a_j^+ a_j - t_{\text{coh}}(\omega) \sum_j (a_j^+ a_{j+1} + \text{h.c.}) \\ &= \sum_k \left(\varepsilon_{\text{coh}}(\omega) - 2t_{\text{coh}}(\omega)\cos ka\right) a_k^+ a_k \end{aligned} \quad (1)$$

To apply CPA, which uses the coherent medium to approximate the alloy system, we need to remove the site $j$ in the coherent medium and later re-insert the real monomer or THDBT blocks. We can divide the space from the local site "$j$" part and the rest "environment" excluding site $j$, called $\bar{j}$. $\{\bar{j}\} = 1, 2, ..., j-1, j+1, ...N.$. The inverse CPA Green's function can be written as

$$G_0^{-1}(\omega) = \omega - H_0 = \begin{pmatrix} \omega - \varepsilon_{\text{coh}}(\omega) & T \\ T^+ & \left[G_0^{\bar{j}}(\omega)\right]^{-1} \end{pmatrix} \quad \text{(S4-Eq5)}$$

where the (1,1) block is the Hamiltonian for site $j$, $T$ is a 1 x (N-1) row vector $T = t_{\text{coh}}(\omega)(|j-1\rangle + |j+1\rangle)$ with $i^{\text{th}}$ component $T_i = \langle i|\omega - H_0|j\rangle = t_{\text{coh}}(\omega)(\delta_{i,j-1} + \delta_{i,j+1}), i \in \{\bar{j}\}$, and $G_0^{\bar{j}}(\omega) = \dfrac{1}{\omega I - H_0^{\bar{j}}}$ is the "cavity Green's function" that excludes the site $j$ and other couplings with site $j$.

The local Green's function $G_{0,jj}(\omega)$ at site $j$ can be computed using Schur complement as

$$\begin{aligned} G_{0,jj}(\omega) &= \langle j|G_0(\omega)|j\rangle = \langle j|\dfrac{1}{\omega - H_0}|j\rangle \\ &= \left(\omega - \varepsilon_{\text{coh}}(\omega) - T G_0^{\bar{j}}(\omega) T^+\right)^{-1} \\ &= \dfrac{1}{\omega - \varepsilon_{\text{coh}}(\omega) - t_{\text{coh}}^2(\omega) R(\omega)} \end{aligned} \quad \text{(S4-Eq6)}$$

in which

$$R(\omega) = \langle j-1|G_0^{\bar{j}}(\omega)|j-1\rangle + \langle j+1|G_0^{\bar{j}}(\omega)|j+1\rangle$$

$$= \frac{a}{2\pi}\int dk \frac{\sin^2 ka}{\omega - \varepsilon_{coh}(\omega) + 2t_{coh}\cos ka}$$

$$= \frac{(\omega - \varepsilon_{coh}) - \sqrt{(\omega - \varepsilon_{coh})^2 - 4t_{coh}^2}}{t_{coh}^2}$$
(S4-Eq7)

is the "cavity return," aka the environmental propagator "seen" when going one step away and return to site $j$. The term $t_{coh}^2(\omega)R(\omega)$ is the local self-energy to quantify how the "environment" (rest of the system excluding $j$) can impact the site $j$ in effective medium.

As a sanity check, with translational invariance, we can also first obtain the energy-momentum space Green's function

$$G_0(k,\omega) = \frac{1}{\omega - \varepsilon_{coh}(\omega) + 2t_{coh}(\omega)\cos ka + i0^+}$$
(S4-Eq8)

After Fourier transform, we have the real-space, on-site Green's function written as

$$G_{0,jj}(\omega) = \frac{a}{2\pi}\int G_0(k,\omega)dk = \frac{a}{2\pi}\int_{-\pi/a}^{+\pi/a} \frac{1}{\omega - \varepsilon_{coh}(\omega) + 2t_{coh}(\omega)\cos ka + i0^+} dk$$

$$= \frac{1}{\sqrt{(\omega - \varepsilon_{coh}(\omega))^2 - 4t_{coh}^2(\omega)}}$$
(S4-Eq9)

which gives the same results as Eqs. *(S4-Eq6) and (S4-Eq7)* before. The CPA local density-of-states can then be written as

$$\rho(\omega) = -\frac{1}{\pi}\text{Im}\, G_{0,jj}(\omega + i0^+)$$
(S4-Eq10)

Now we replace the coherent background at local site $j$ with actual monomer A or filler B, and the environment stays the same. The local Green's function, when ensuring single-site locality, can be written as

$$G_{\sigma,jj}(\omega) = \frac{1}{\omega - \varepsilon_\sigma(\omega) - t_{\sigma\sigma}^2 R(\omega)}, \sigma = A, B$$
(S4-Eq11)

Then, defining the single-site effective scattering potential vertex at site A or B as,

$$V_\sigma(\omega) = \varepsilon_\sigma - \varepsilon_{coh}(\omega) + \left(t_{\sigma\sigma}^2 - t_{coh}^2(\omega)\right)R(\omega), \sigma = A, B$$
(S4-Eq12)

we can write down the Dyson's equation

$$G_{\sigma,jj}(\omega) = G_{0,jj}(\omega) + G_{0,jj}(\omega)V_\sigma(\omega)G_{\sigma,jj}(\omega) \tag{S4-Eq13}$$

Similarly, the single-site T-matrix can be written as

$$T_A(\omega) = \frac{V_A(\omega)}{1 - V_A(\omega)G_{0,jj}(\omega)}$$

$$T_B(\omega) = \frac{V_B(\omega)}{1 - V_B(\omega)G_{0,jj}(\omega)} \tag{S4-Eq14}$$

Finally, we can impose the CPA self-consistent condition, at each frequency,

$$(1-c)T_A(\omega) + cT_B(\omega) = 0 \tag{S4-Eq15}$$

Physically, this means that replacing the coherent effective medium to real local site does not lead to additional scattering. Besides vanishing site-scattering, we also have the bonding scattering condition, i.e.,

$$(1-c)\frac{t_{AA}^2 - t_{coh}^2(\omega)}{1 - V_A(\omega)G_{0,jj}(\omega)} + c\frac{t_{BB}^2 - t_{coh}^2(\omega)}{1 - V_B(\omega)G_{0,jj}(\omega)} = 0 \tag{S4-Eq16}$$

if we subtract Eq. (16) from(15), we have:

$$(1-c)\frac{\varepsilon_A - \varepsilon_{coh}(\omega)}{1 - V_A(\omega)G_{0,jj}(\omega)} + c\frac{\varepsilon_B - \varepsilon_{coh}(\omega)}{1 - V_B(\omega)G_{0,jj}(\omega)} = 0 \tag{S4-Eq17}$$

which is the average energy condition, highlighting that the effective coherent medium does not change the averaged energy.

**Section 4.3 Heat capacity and thermal conductivity.**

Since the trend of thermal conductivity and specific heat capacity are the same as a function of filler fraction, according to the recent work,[2]

$$C_V(T) = \int_0^{+\infty} d\omega \frac{\omega^2 e^{\beta\omega}}{T^2(e^{\beta\omega}-1)^2}\rho(\omega) \tag{S4-Eq18}$$

Similarly, we can do thermal conductivity,

$$k(T) = \frac{1}{T}\int_0^{+\infty} d\omega \omega^2 \left(-\frac{\partial n_B}{\partial \omega}\right)\frac{a}{2\pi}\int_{-\pi/a}^{+\pi/a} dk v_k^2(\omega)[A_0(k,\omega)]^2$$

$$v_k(\omega) = \partial_k \left(\varepsilon_{coh}(\omega) - 2t_{coh}(\omega)\cos ka\right) = 2a\sin ka \tag{S4-Eq19}$$

in which $A_0(k,\omega) = -\frac{1}{\pi}\operatorname{Im} G_0(k,\omega)$ is the spectral function of the CPA Green's function Eq. *(S4-Eq8)*.

We also summarize the brief procedure to use CPA to compute the thermal conductivity in a disordered polymer:

1) Obtain energies from MD/DFT calculation, including $\varepsilon_A$, $\varepsilon_B$, $t_{AA}$, and $t_{BB}$ as fitting parameters. Set $t_{AB} = \sqrt{t_{AA} t_{BB}}$.

2) Pick a filler fractional concentration $0 < c < 1$, substitute back to the BEB-CPA equations *Eq. (S4-Eq15) and Eq. (S4-Eq17)*, to obtain the CPA parameters $\varepsilon_{coh}(\omega)$ and $t_{coh}(\omega)$.

3) Use *Eq. (S4-Eq9)* to compute CPA local DOS $\rho(\omega)$, and use *Eq. (S4-Eq8)* to compute CPA spectral function $A_0(k,\omega)$.

4) Use *Eq. (S4-Eq18) and (S4-Eq19)* to compute heat capacity and thermal conductivity.

**Figures. S1 to S10**

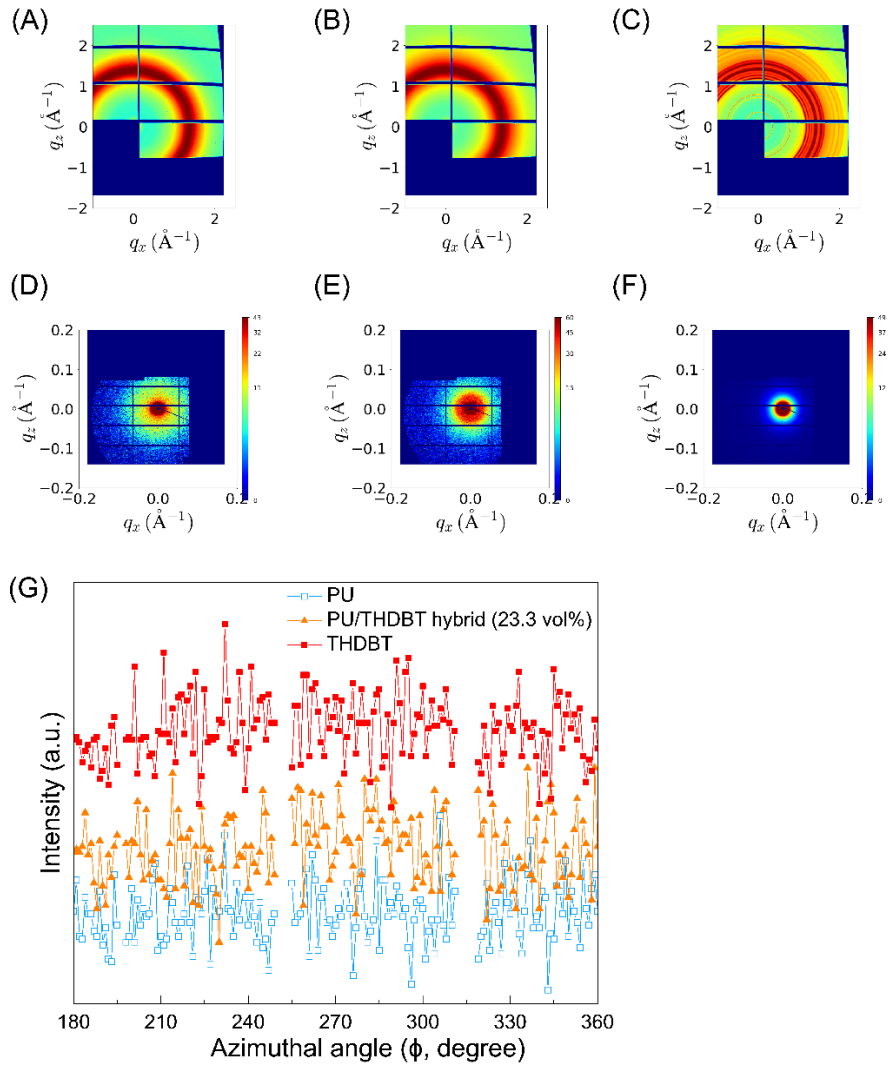

**Figure. S1. Structural characterizations using synchrotron X-ray scattering. (A)** Wide-angle X-ray scattering patterns of pure PU thin film. **(B)** Wide-angle X-ray scattering patterns of PU/THDBT hybrid ($23.3\ vol\%$) thin film. **(C)** Wide-angle X-ray scattering patterns of compressed pellets of THDBT fillers. **(D)** Small-angle X-ray scattering patterns of pure PU thin film. **(E)** Small-angle X-ray scattering patterns of pure PU thin film PU/THDBT hybrid ($23.3\ vol\%$) thin film. **(F)** Small-angle X-ray scattering patterns of compressed pellets of THDBT fillers. **(G)** Azimuthal line cuts integrated around the peak ($q\ =\ 1.374\ \text{Å}^{-1}$) within a $\pm 0.006\ \text{Å}^{-1}$ width of the PU in pure PU film. Azimuthal line cuts integrated around the peak ($q\ =\ 1.400\ \text{Å}^{-1}$) within a $\pm 0.006\ \text{Å}^{-1}$ width of the THDBT in PU/THDBT hybrid ($23.3\ vol\%$). Azimuthal line cuts integrated around the peak ($q\ =\ 1.400\ \text{Å}^{-1}$) within a $\pm 0.006\ \text{Å}^{-1}$ width of the THDBT in compressed pellets of THDBT fillers.

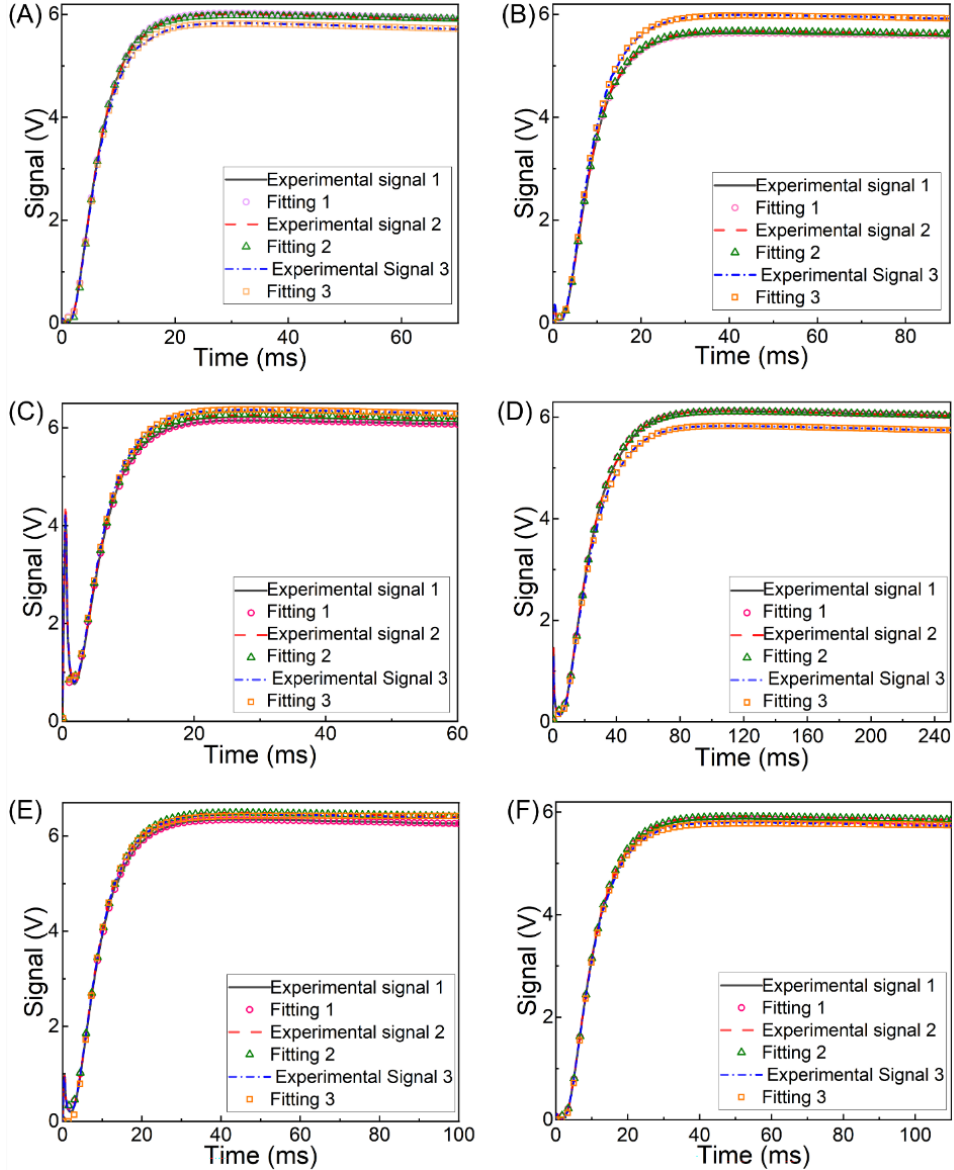

**Figure S2. Measured cross-plane thermal diffusivities of pure PU thin films.** In order to minimize random errors and ensure the reproducibility of thermal diffusivity experimental measurements made using the laser flash method, each sample was tested three times. Thermal diffusivity results of 6 different PU thin films are shown. All samples were coated with graphite spray (DGF 123) on both sides before thermal diffusivity testing. The "transparent" model in the LFA 467 software was used to fit the crossplane thermal diffusivity experimental signals obtained from the laser flash method. **(A)** Experimental and fitting results for cross-plane thermal diffusivities of thin film (PU) with a thickness of $0.0547\ mm$. **(B)** Experimental and fitting results for cross-plane thermal diffusivities of thin film (PU) with a thickness of $0.0680\ mm$. **(C)** Experimental and fitting results for cross-plane thermal diffusivities of thin film (PU) with a thickness of $0.0590\ mm$. **(D)** Experimental and fitting results for cross-plane thermal diffusivities of thin film (PU) with a thickness of $0.1140\ mm$. **(E)** Experimental and fitting results for cross-plane thermal diffusivities of thin film (PU) with a thickness of $0.0665\ mm$. **(F)** Experimental and fitting results for cross-plane thermal diffusivities of thin film (PU) with a thickness of $0.0720\ mm$.

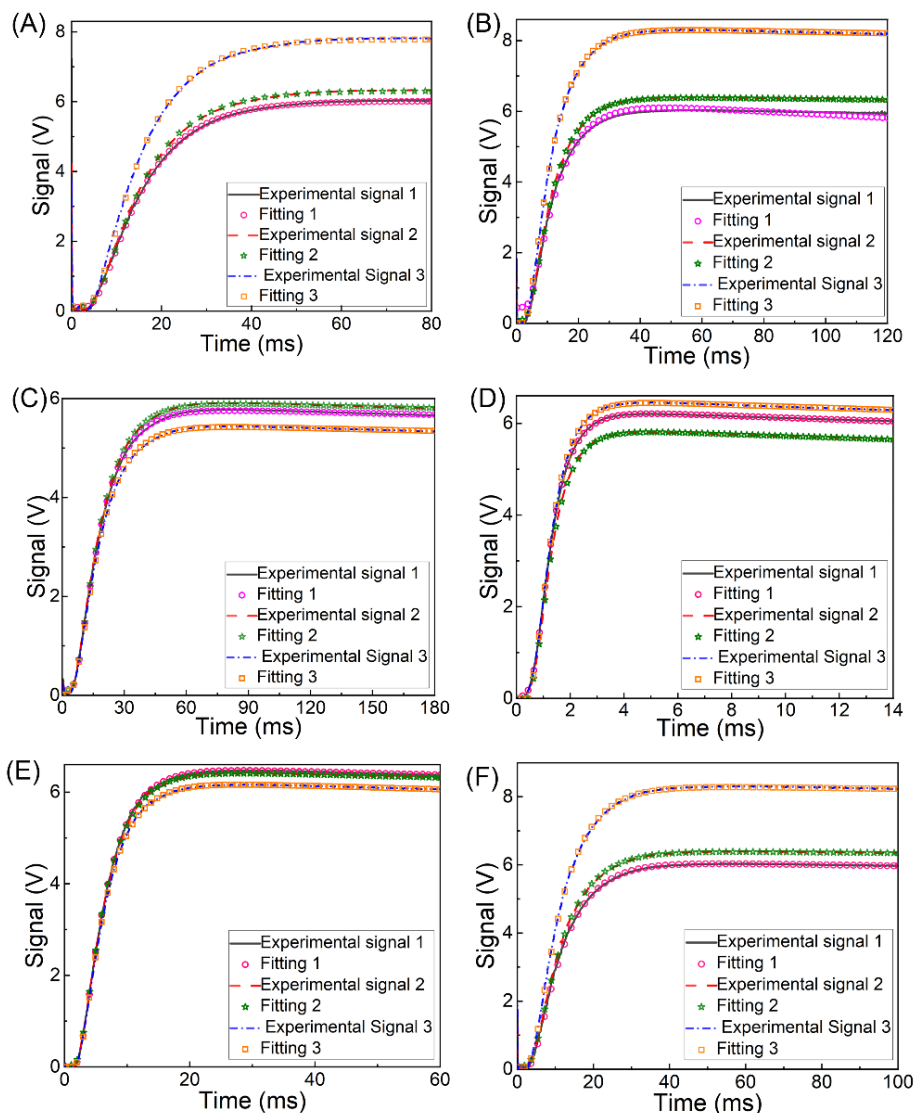

**Figure S3. Measured cross-plane thermal diffusivities of PU/THDBT hybrids (4.5 vol%).** In order to minimize random errors and ensure the reproducibility of thermal diffusivity experimental measurements made using the laser flash method, each sample was tested three times. Thermal diffusivity results of 6 different PU/THDBT hybrids (4.5 $vol\%$) are shown. All samples were coated with graphite spray (DGF 123) on both sides before thermal diffusivity testing. The "transparent" model in the LFA 467 software was used to fit the cross-plane thermal diffusivity experimental signals obtained from the laser flash method. **(A)** Experimental and fitting results for cross-plane thermal diffusivities of thin film (PU/THDBT hybrids (4.5 $vol\%$)) with a thickness of $0.0715\ mm$. **(B)** Experimental and fitting results for cross-plane thermal diffusivities of thin film (PU/THDBT hybrids (4.5 $vol\%$) with a thickness of $0.0750\ mm$. **(C)** Experimental and fitting results for cross-plane thermal diffusivities of thin film(PU/THDBT hybrids (4.5 $vol\%$)) with a thickness of $0.0940\ mm$. **(D)** Experimental and fitting results for cross-plane thermal diffusivities of thin film (PU/THDBT hybrids (4.5 vol%) with a thickness of $0.0225\ mm$. **(E)** Experimental and fitting results for cross-plane thermal diffusivities of thin film (PU/THDBT hybrids (4.5 $vol\%$)) with a thickness of $0.0535\ mm$. **(F)** Experimental and fitting results for cross-plane thermal diffusivities of thin film (PU/THDBT hybrids (4.5 $vol\%$)) with a thickness of $0.0850\ mm$.

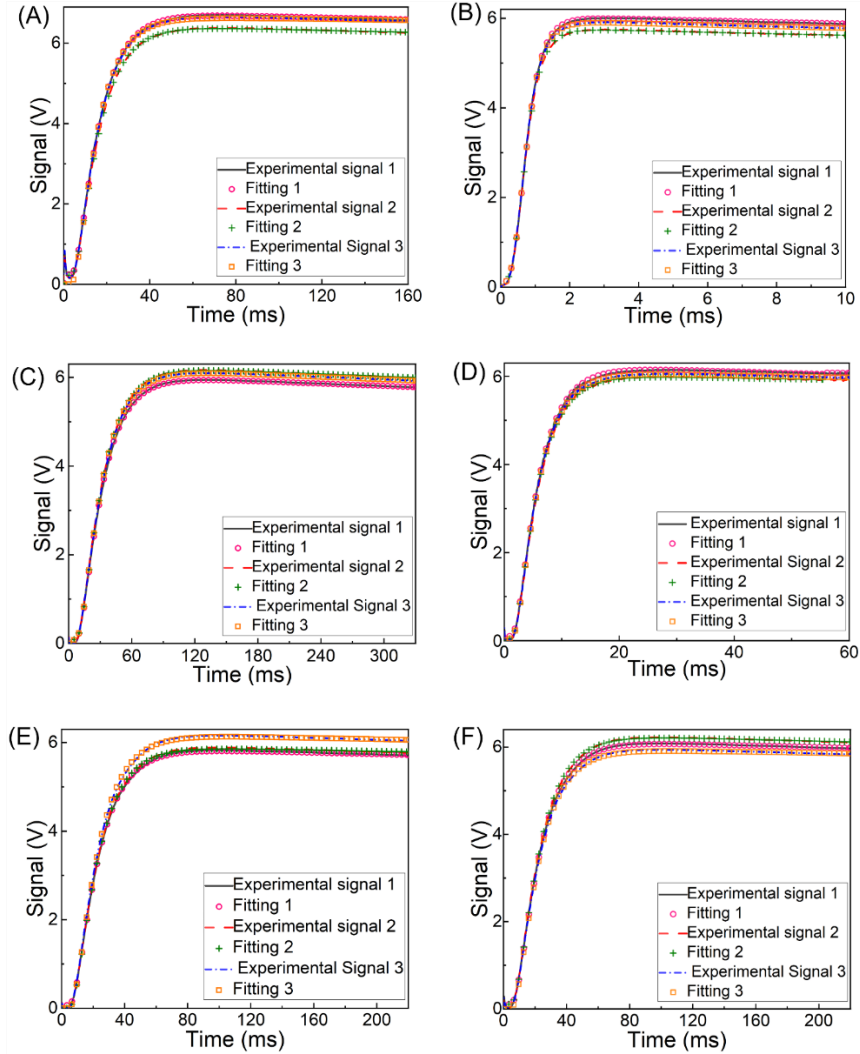

**Figure. S4. Measured cross-plane thermal diffusivities of PU/THDBT hybrids (13.9 $vol\%$).** In order to minimize random errors and ensure the reproducibility of thermal diffusivity experimental measurements made using the laser flash method, each sample was tested three times. Thermal diffusivity results of 6 different PU/THDBT 13.9 $vol\%$ hybrid samples are shown. All samples were coated with graphite spray (DGF 123) on both sides before thermal diffusivity testing. The "transparent" model in the LFA 467 software was used to fit the crossplane thermal diffusivity experimental signals obtained from the laser flash method. **(A)** Experimental and fitting results for cross-plane thermal diffusivities of thin film (PU/THDBT hybrids (13.9 $vol\%$)) with a thickness of $0.0870\ mm$. **(B)** Experimental and fitting results for cross-plane thermal diffusivities of thin film (PU/THDBT hybrids (13.9 $vol\%$)) with a thickness of $0.1320\ mm$. **(C)** Experimental and fitting results for cross-plane thermal diffusivities of thin (PU/THDBT hybrids (13.9 vol%)) with a thickness of $0.0560\ mm$. **(D)** Experimental and fitting results for cross-plane thermal diffusivities of thin film (PU/THDBT hybrids (13.9 $vol\%$)) with a thickness of $0.1050\ mm$. **(E)** Experimental and fitting results for cross-plane thermal diffusivities of thin film (PU THDBT Hybrid (PU/THDBT hybrids (13.9 $vol\%$)) with a thickness of $0.1030\ mm$. **(F)** Experimental and fitting results for cross-plane thermal diffusivities of thin film (PU/THDBT hybrids (13.9 $vol\%$)) with a thickness of $0.1040\ mm$.

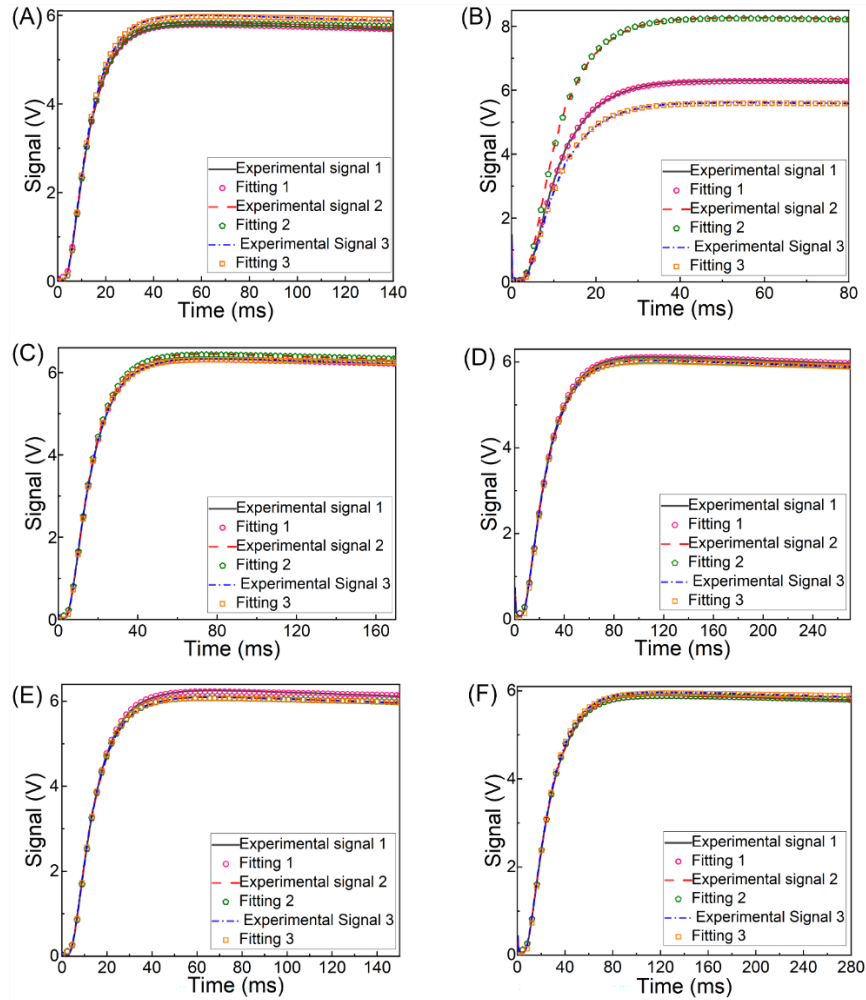

**Figure. S5. Measured cross-plane thermal diffusivities of PU/THDBT hybrids (23.3 $vol\%$).** In order to minimize random errors and ensure the reproducibility of thermal diffusivity experimental measurements made using the laser flash method, each sample was tested three times. Thermal diffusivity results of 6 different PU/THDBT 23.3 vol% hybrid samples are shown. All samples were coated with graphite spray (DGF 123) on both sides before thermal diffusivity testing. The "transparent" model in the LFA 467 software was used to fit the crossplane thermal diffusivity experimental signals obtained from the laser flash method. **(A)** Experimental and fitting results for cross-plane thermal diffusivities of thin film (PU/THDBT hybrids (23.3 $vol\%$)) with a thickness of 0.0800 $mm$. **(B)** Experimental and fitting results for cross-plane thermal diffusivities of thin film (PU/THDBT hybrids (23.3 $vol\%$)) with a thickness of 0.0770 $mm$. **(C)** Experimental and fitting results for cross-plane thermal diffusivities of thin film (PU/THDBT hybrids (23.3 $vol\%$)) with a thickness of 0.0870 $mm$. **(D)** Experimental and fitting results for cross-plane thermal diffusivities of thin film (PU/THDBT hybrids (23.3 $vol\%$)) with a thickness of 0.1160 $mm$. **(E)** Experimental and fitting results for cross-plane thermal diffusivities of thin film (PU/THDBT hybrids (23.3 $vol\%$)) with a thickness of 0.0810 $mm$. **(F)** Experimental and fitting results for cross-plane thermal diffusivities of thin film (PU/THDBT hybrids (23.3 $vol\%$)) with a thickness of 0.1165 $mm$.

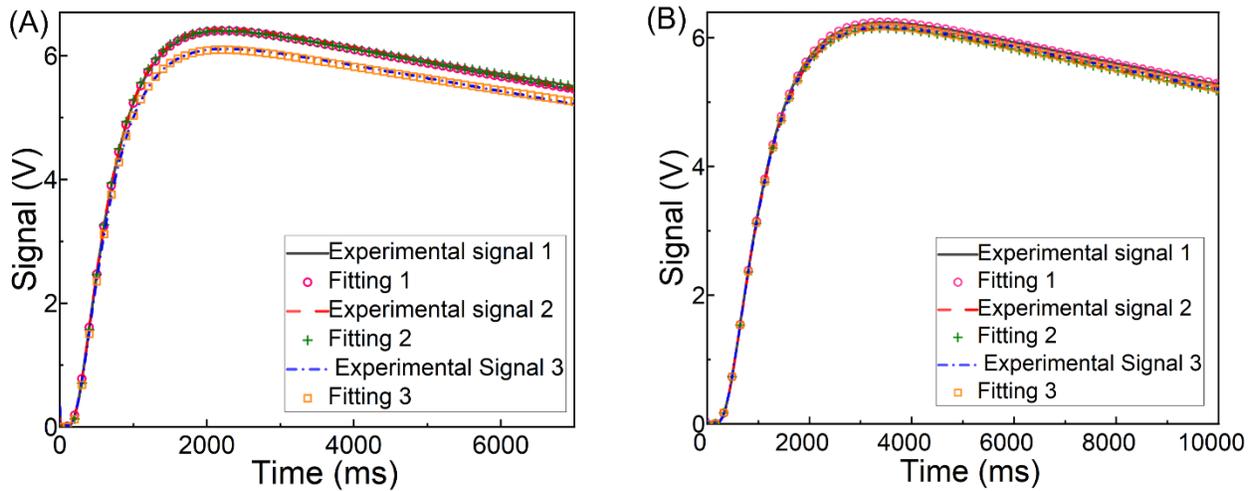

**Figure S6. Measured cross-plane thermal diffusivities of compressed pellets of THDBT fillers.** In order to minimize random errors and ensure the reproducibility of the thermal diffusivity experimental measurements made using the laser flash method, each sample was tested three times. Thermal diffusivity results of 2 different THDBT samples are shown. All samples were coated with graphite spray (DGF 123) on both sides before thermal diffusivity testing. The "penetration" model in the LFA 467 software was used to fit the crossplane thermal diffusivity experimental signals obtained from the laser flash method. **(A)** Experimental and fitting results for cross-plane thermal diffusivities of compressed pellets of THDBT fillers with thickness of $1.1550\ mm$. **(B)** Experimental and fitting results for cross-plane thermal diffusivities of compressed pellets of THDBT fillers with thickness of $0.9560\ mm$.

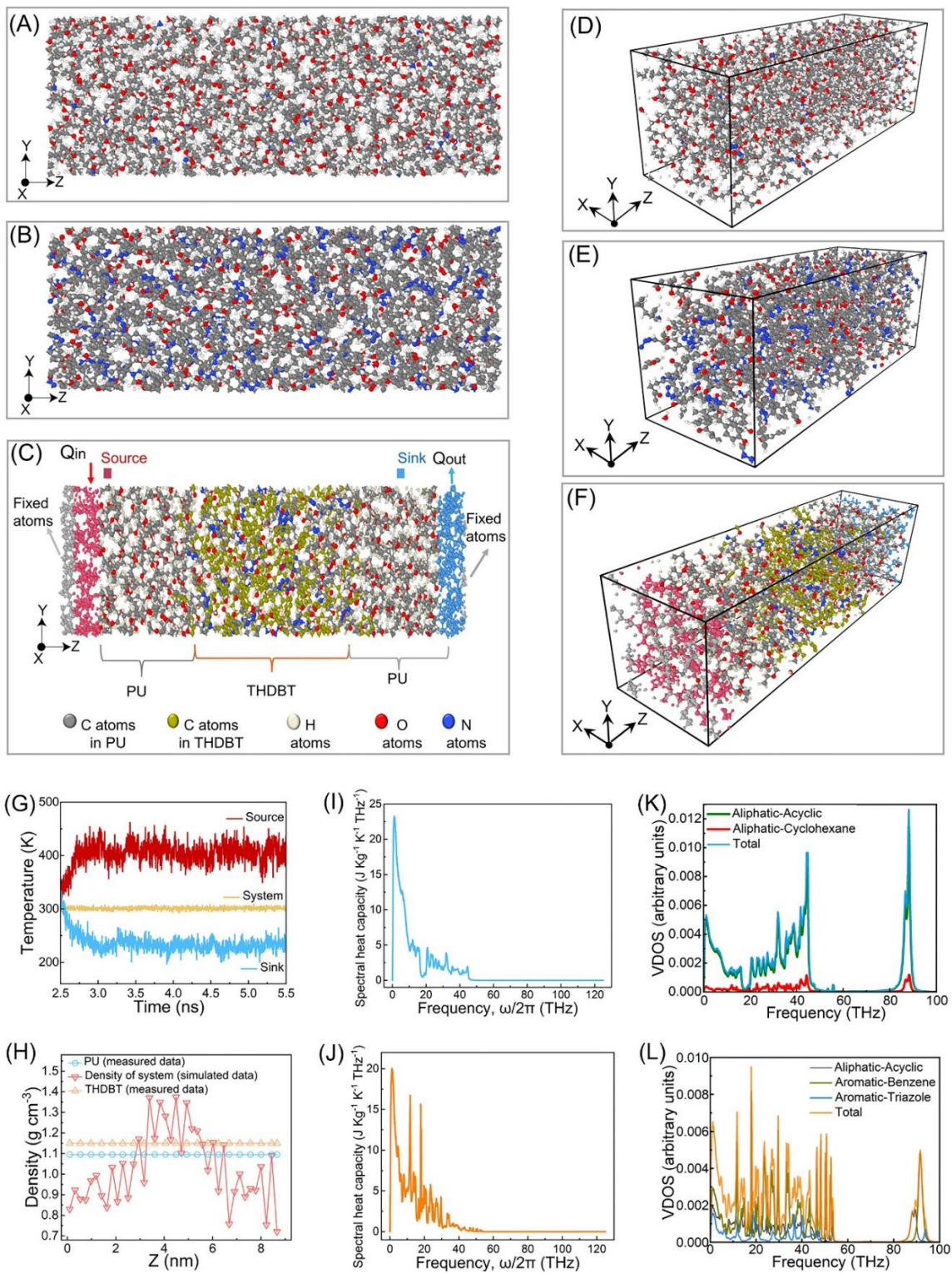

**Figure S7**. **(A)** Simulation domain of PU. **(B)** Simulation domain of THDBT. **(C)** Simulation domain of PU/THDBT interface. **(D)** Three-dimensional view of simulation domain of PU. **(E)** 3-dimensional view of simulation domain of THDBT. **(F)** Three-dimensional view of simulation domain of PU/THDBT interface. **(G)** Temperature of heat source, heat sink and overall

temperature of the system with time. The average temperature of the heat source and sink becomes steady after 3.5 ns. **(H)** Density of the system at different positions along the z axis. **(I)** Spectral heat capacity of PU **(J)** Spectral heat capacity of THDBT. **(K)** the vibrational density of states (VDOS) contributions from different chain segments of PU. **(L)** the vibrational density of states (VDOS) contributions from different functional groups of THDBT.

For clarity, in Figures S7K and S7L, the polymer backbones are divided into four segments based on chemical configuration: (i) Aromatic benzene, consisting of C and H atoms in the benzene ring; (ii) Aromatic triazole, consisting of C, H, and N atoms in the triazole ring, (iii) Aliphatic acyclic backbone, consisting of C, H, and O atoms; and (iv) Aliphatic cyclic backbone, consisting of C, H of cyclohexene. A distinct contrast is observed in the $45 - 55\ THz$ frequency range, where THDBT exhibits pronounced peaks that are absent in PU. In this region, the dominant contribution arises from the benzene ring vibrations of THDBT. The presence of these high-frequency phonon modes shifts the overall VDOS of THDBT toward higher frequencies relative to PU. Consequently, the phonon population in the $20 - 40\ THz$ range is reduced. This results in a lower specific heat capacity for THDBT compared to PU As only lower frequency modes ($< 45\ THz$) contribute to heat capacity.

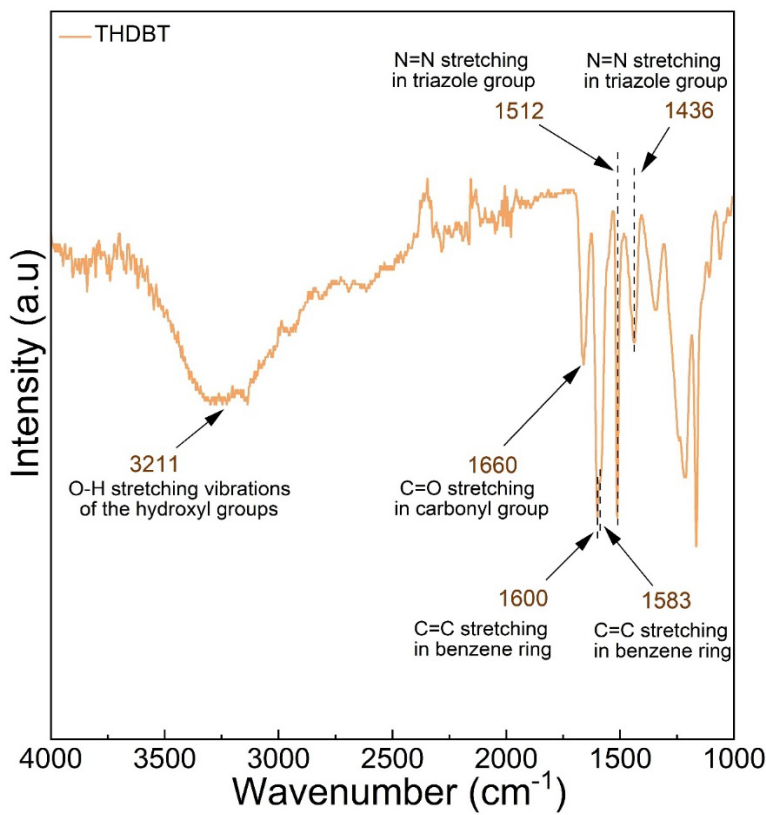

**Figure S8.** ATR-FTIR spectra of compressed pellets made from THDBT. Please note that ATR-FTIR spectrum for THDBT in Figure 6D, because the O–H stretching vibrations in pure THDBT are much weaker than the N–H stretching vibrations in pure PU, the O–H peaks ($\sim 3211\ cm^{-1}$) appears flattened when all ATR-FTIR spectra are plotted together. For clarity, the individual ATR-FTIR spectrum of pure THDBT is shown in this Figure S8.

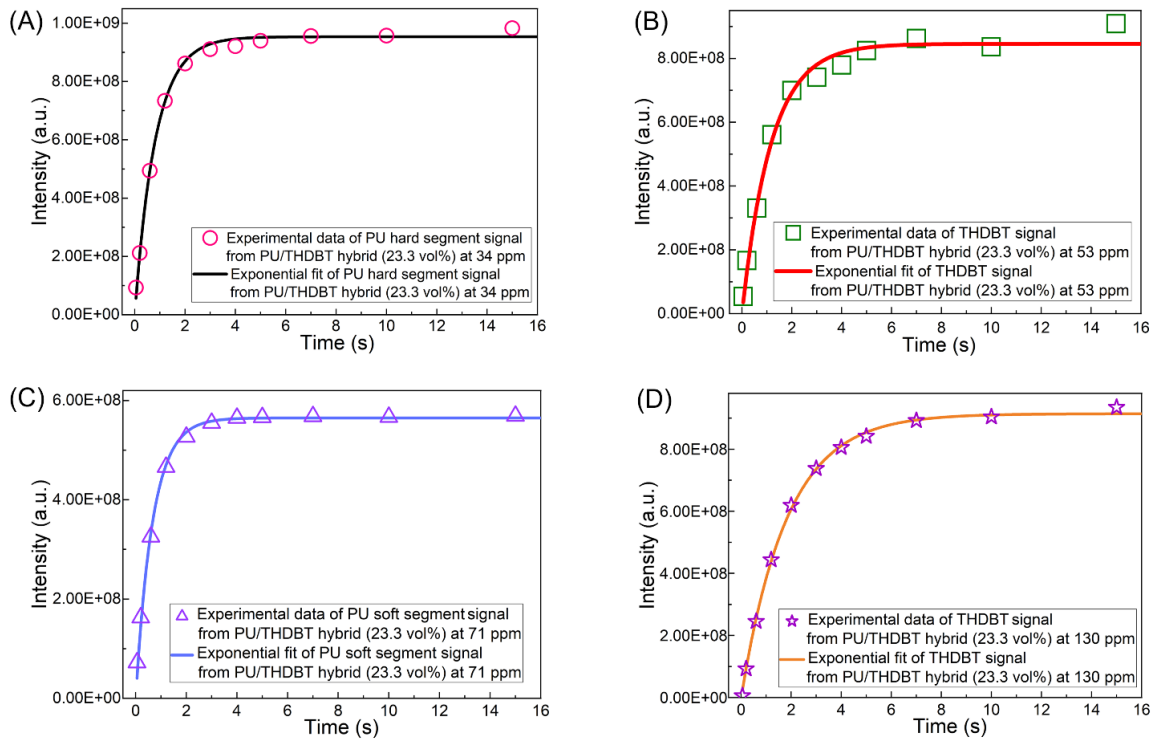

**Figure S9.** $^1$H T1 recovery curves for PU/THDBT hybrid (23.3 *vol*%). Intensity is shown in arbitrary units (a.u.). (A) Polyurethane (PU) hard segment signal at 34 ppm from PU/THDBT hybrid (23.3 *vol*%). (B) THDBT signal at 53 *ppm* from PU/THDBT hybrid (23.3 *vol*%). (C) PU soft segment signal at 71 *ppm* from PU/THDBT hybrid (23.3 *vol*%). (D) THDBT signal at 130 *ppm* from PU/THDBT hybrid (23.3 *vol*%).

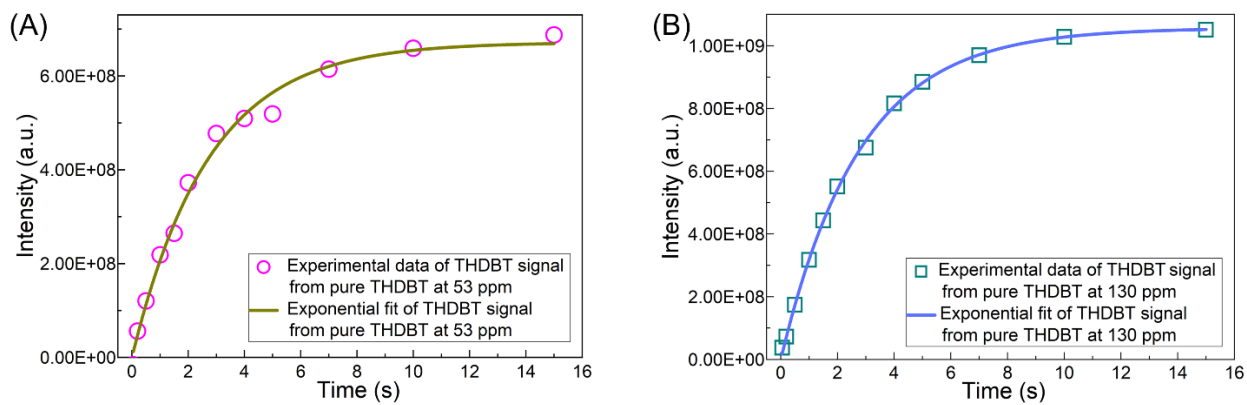

**Figure. S10.** $^1$H T1 recovery curves for pure THDBT. Intensity is shown in arbitrary units (a.u.). **(A)** THDBT signal at 53 *ppm* from pure THDBT. **(B)** THDBT signal at 130 *ppm* from pure THDBT.

**Table S1**. $^1$H T1 relaxation times (s).

|  | PU Soft segments | PU hard segments | THDBT |
|---|---|---|---|
| PU/THDBT hybrid films with a THDBT volume fraction of 23.3 *vol*% | $0.67 \pm 0.03\ (s)$ | $0.82 \pm 0.03\ (s)$ | $1.83 \pm 0.04\ (s)$ |
| THDBT | not applicable | not applicable | $2.78 \pm 0.05\ (s)$ |